\documentclass[twocolumn,astrosymb]{aastex7}

\usepackage{graphicx, caption, subcaption, rotating}
\usepackage[utf8]{inputenc}
\usepackage{newunicodechar}
\newunicodechar{−}{\ensuremath{-}}
\usepackage{graphicx, caption, subcaption, rotating}
\DeclareCaptionFormat{cont}{#1 (cont.)#2#3\par}

\shorttitle{Molecular Gas in Major Mergers Hosting Dual and Single AGN}
\shortauthors{Johnstone et al.}

\begin{document}

\title{Molecular Gas in Major Mergers Hosting Dual and Single AGN at $<$10 kpc Nuclear Separations}

\correspondingauthor{Makoto A. Johnstone}
\email{fhh3kp@virginia.edu}

\author[0000-0001-7690-3976]{Makoto A. Johnstone}
\affiliation{Department of Astronomy, University of Virginia, 530 McCormick Road, Charlottesville, VA 22903, USA}
\affil{Instituto de Astrof{\'{\i}}sica, Facultad de F{\'{i}}sica, Pontificia Universidad Cat{\'{o}}lica de Chile, Campus San Joaqu{\'{\i}}n, Av. Vicu{\~{n}}a Mackenna 4860, Macul Santiago, Chile, 7820436} 
\email{fhh3kp@virginia.edu}

\author[0000-0001-7568-6412]{Ezequiel Treister}
\affil{Instituto de Alta Investigaci{\'{o}}n, Universidad de Tarapac{\'{a}}, Casilla 7D, Arica, Chile}
\email{etreister@academicos.uta.cl}

\author[0000-0002-8686-8737]{Franz E. Bauer}
\affiliation{Instituto de Alta Investigaci{\'{o}}n, Universidad de Tarapac{\'{a}}, Casilla 7D, Arica, Chile}
\email{franz.e.bauer@gmail.com}

\author[0000-0001-9910-3234]{Chin-Shin Chang}
\affiliation{Joint ALMA Observatory, Avenida Alonso de Cordova 3107, Vitacura 7630355, Santiago, Chile}
\email{ChinShin.Chang@alma.cl}

\author[0000-0003-0522-6941]{Claudia Cicone}
\affiliation{Institute of Theoretical Astrophysics, University of Oslo, Postboks 1029, NO-0315 Oslo, Norway}
\email{claudia.cicone@astro.uio.no}

\author[0000-0002-7998-9581]{Michael J. Koss}
\affil{Eureka Scientific, 2452 Delmer Street Suite 100, Oakland, CA 94602-3017, USA}
\email{mike.koss@eurekasci.com}

\author[0000-0001-8931-1152]{Ignacio del Moral-Castro}
\affil{Instituto de Astrof{\'{\i}}sica, Facultad de F{\'{i}}sica, Pontificia Universidad Cat{\'{o}}lica de Chile, Campus San Joaqu{\'{\i}}n, Av. Vicu{\~{n}}a Mackenna 4860, Macul Santiago, Chile, 7820436} \email{ignaciodelmoralcastro.astro@gmail.com}

\author[0000-0002-2713-0628]{Francisco Muller-Sanchez}
\affiliation{Department of Physics and Materials Science, The University of Memphis, 3720 Alumni Avenue, Memphis, TN 38152, USA}
\email{F.Muller.Sanchez@memphis.edu}

\author[0000-0003-3474-1125]{George C. Privon}
\affiliation{National Radio Astronomy Observatory, 520 Edgemont Road, Charlottesville, VA 22903, USA}
\affiliation{Department of Astronomy, University of Florida, P.O. Box 112055, Gainesville, FL 32611, USA}
\affiliation{Department of Astronomy, University of Virginia, 530 McCormick Road, Charlottesville, VA 22903, USA}
\email{gprivon@nrao.edu}

\author[0000-0001-5231-2645]{Claudio Ricci}
\affiliation{Instituto de Estudios Astrofísicos, Facultad de Ingeniería y Ciencias, Universidad Diego Portales, Avenida Ejercito Libertador 441, Santiago, Chile}
\affiliation{Kavli Institute for Astronomy and Astrophysics, Peking University, Beijing 100871, People’s Republic of China}
\email{claudio.ricci@mail.udp.cl}

\author[0000-0002-0438-3323]{Nick Scoville}
\affiliation{Cahill Center for Astrophysics, California Institute of Technology, 1216 East California Boulevard, Pasadena, CA 91125, USA}
\email{nzs@astro.caltech.edu}

\author[0000-0001-8349-3055]{Giacomo Venturi}
\affiliation{Scuola Normale Superiore, Piazza dei Cavalieri 7, 56126 Pisa, Italy}
\affiliation{INAF – Osservatorio Astrofisico di Arcetri, Largo E. Fermi 5, 50125 Firenze, Italy}
\email{giacomo.venturi1@sns.it}

\author[0000-0003-0057-8892]{Loreto Barcos-Munoz}
\affiliation{National Radio Astronomy Observatory, 520 Edgemont Road, Charlottesville, VA 22903, USA}
\email{lbarcos@nrao.edu}

\author[0000-0003-3498-2973]{Lee Armus}
\affiliation{IPAC, California Institute of Technology, 1200 East California Boulevard, Pasadena, CA 91125, USA}
\email{lee@ipac.caltech.edu}

\author[0000-0002-2183-1087]{Laura Blecha}
\affiliation{Department of Physics, University of Florida, 2001 Museum Rd., Gainesville, FL 32611, USA}
\email{lblecha@ufl.edu}

\author[0000-0002-0930-6466]{Caitlin Casey}
\affiliation{Department of Physics, University of California, Santa Barbara, Santa Barbara, CA 93106, USA}
\affiliation{Cosmic Dawn Center (DAWN), Denmark}
\email{cmcasey@ucsb.edu}

\author[0000-0001-8627-4907]{Julia Comerford}
\affiliation{Department of Astrophysical and Planetary Sciences, University of Colorado, Boulder, CO 80309, USA}
\email{julie.comerford@colorado.edu}

\author[0000-0003-2638-1334]{Aaron Evans}
\affiliation{Department of Astronomy, University of Virginia, 530 McCormick Road, Charlottesville, VA 22903, USA}
\affiliation{National Radio Astronomy Observatory, 520 Edgemont Road, Charlottesville, VA 22903, USA}
\email{aevans@virginia.edu}

\author[0000-0002-6808-2052]{Taiki Kawamuro}
\affiliation{RIKEN Cluster for Pioneering Research, 2-1 Hirosawa, Wako, Saitama 351-0198, Japan}
\email{taiki.kawa15@gmail.com}

\author[0000-0001-7421-2944]{Anne M. Medling}
\affiliation{Department of Physics \& Astronomy and Ritter Astrophysical Research Center, University of Toledo, Toledo, OH 43606, USA}
\affiliation{ARC Centre of Excellence for All Sky Astrophysics in 3 Dimensions (ASTRO 3D), Australia}
\email{anne.medling@utoledo.edu}

\author[0000-0002-2985-7994]{Hugo Messias}
\affiliation{Joint ALMA Observatory, Avenida Alonso de Cordova 3107, Vitacura 7630355, Santiago, Chile}
\email{hugo.messias@alma.cl}

\author[0000-0001-6920-662X]{Neil Nagar}
\affiliation{Universidad de Concepción, Departamento de Astronomía, Casilla 160-C, Concepción, Chile}
\email{nagar@astro-udec.cl}

\author[0000-0003-0006-8681]{Alejandra Rojas}
\affiliation{Departamento de F\'isica, Universidad T\'ecnica Federico Santa Mar\'ia, Campus San Joaquin, Avenida Vicu\~na Mackenna 3939,
San Joaqu\'in, Santiago, Chile}
\email{ale.rojaslilayu@gmail.com}

\author[0000-0002-1233-9998]{David Sanders}
\affiliation{Institute for Astronomy, 2680 Woodlawn Drive, University of Hawaii, Honolulu, HI 96822, USA}
\email{sanders@ifa.hawaii.edu}

\author[0000-0002-3683-7297]{Benny Trakhtenbrot}
\affiliation{School of Physics and Astronomy, Tel Aviv University, Tel Aviv 69978, Israel}
\email{benny.trakht@gmail.com}

\author[0000-0002-1912-0024]{Vivian U}
\affiliation{Department of Physics and Astronomy, 4129 Frederick Reines Hall, University of California, Irvine, CA 92697, USA}
\email{vivianu@uci.edu}

\author[0000-0002-0745-9792]{Meg Urry}
\affiliation{Yale Center for Astronomy \& Astrophysics, Physics Department, New Haven, CT 06520, USA}
\email{meg.urry@yale.edu}

\begin{abstract}
We present high-resolution ($\sim$\,50$-$100 pc) Atacama Large Millimeter Array (ALMA) observations of $^{12}$CO(2-1) or $^{12}$CO(1-0) emission in seven local ($z$ $\lesssim$ 0.05) major mergers --- five of which are dual active galactic nuclei (AGN) systems, and two of which are single AGN systems. We model the molecular gas kinematics through rotating disk profiles using a Bayesian Markov chain Monte Carlo approach. The residuals were then used to isolate non-rotating components of the molecular gas-- the most likely contributor to future SMBH growth. We find that more massive SMBHs have higher surface densities of non-rotating molecular gas within their sphere of influence. This potential molecular gas supply, however, does not correlate with the current accretion efficiency of the SMBHs, suggesting that only a fraction of the observed non-rotating gas is currently reaching the SMBH. Finally, we tentatively find no significant differences in the nuclear molecular gas masses of single AGN and dual AGN hosts, both within the SMBH sphere of influence and within the central kiloparsec. Our results indicate that the probability of occurrence of the dual AGN phenomenon is likely dependent on AGN variability and/or obscuration rather than the availability of molecular gas in the nuclear regions. 

% We find that the accretion efficiency of SMBHs does not depend on the supply of non-rotating molecular gas.  Instead, the nuclear gas structure is likely regulated by the mass of the SMBH itself, with more massive SMBHs having higher surface densities of non-rotating gas within the sphere of influence. Finally, we find no significant differences in the molecular gas properties of single AGN and dual AGN. Our results suggest that the probability of occurrence of the dual AGN phenomenon is dependent on AGN variability and/or obscuration rather than the availability of molecular gas in the nuclear regions. 

\end{abstract}


\section{Introduction} 
\label{sec:intro}

Gas-rich major galaxy mergers are considered to play a fundamental role in
galaxy evolution \citep{sanders88}. These galaxy-galaxy dynamical interactions cause gas to lose angular momentum and to fall into the center of each nucleus \citep[e.g.,][]{barnes91}. Observational evidence collected over decades has confirmed the presence of these large nuclear (dense) molecular gas reservoirs in major galaxy mergers \citep[e.g.,][]{young86,sanders91,young91,combes94,taniguchi98,gao04}. Both observations \citep{ellison08,juneau09,ellison13} and numerical computations \citep[e.g.,][]{barnes91,mihos96,moreno19} indicate that the accumulation of this gas can drive the high star formation rates observed in these systems.

In turn, these high nuclear gas concentrations can also generate significant supermassive black hole (SMBH) growth episodes, resulting in an Active Galactic 
Nucleus (AGN) \citep[e.g.,][]{hopkins06}. While most AGN activity ($\sim$\,90\% of detected AGN by number; \citealt{treister12}) is triggered by internal, secular processes and minor galaxy 
mergers, major mergers are directly linked to the most dramatic SMBH growth events which can be responsible for about half of the total SMBH accretion across cosmic history \citep{treister12}. Several studies have confirmed this conclusion independently (e.g., \citealp{glikman15,kocevski15,donley18,weigel18, Euclid2025}). The large amounts of energy released by star formation and AGN activity can have major implications for cosmological galaxy evolution and the overall accretion history of SMBHs \citep[e.g.,][]{di-matteo08}.

Major galaxy mergers have typical durations of $\sim$10$^9$ years \citep{hopkins08, Lotz2008, moreno19}. During this process, the participating galaxies undergo several stages that can be characterized by their nuclear separations --- ranging from two isolated galaxies ($\sim$\,100s\,kpc) to coalescence ($\leq$\,100s\,pc) \citep{stierwalt13, Ricci2021}. It is also expected that the characteristic timescale of the active SMBH growth phases would be significantly shorter ($\sim$\,10$^5-10^8$\,yrs; \citealt{marconi04,di-matteo05,schawinski15}) than the duration of a galaxy merger. Hence, a fundamental question is: when does most of the associated SMBH growth occur during a major merger, and how much accretion onto the SMBH can be expected within each merger stage? In particular, computational simulations predict that the largest simultaneous star formation and SMBH growth episodes take place when nuclear separations are relatively small ($<$\,10\,kpc; e.g., \citealt{hopkins06,van-wassenhove12}), emitting intense electromagnetic radiation and causing significant effects on the host galaxies \citep{di-matteo08, rupke11,farrah12,veilleux13,cicone14}. 

These effects are most notable when both SMBHs are simultaneously accreting, forming so-called dual AGN systems \citep{rosas-guevara19}. The nuclear regions of these systems are likely subject to high, even Compton-thick, levels of obscuration \citep{blecha18}. Combined with the associated small angular separation, this nuclear obscuration makes dual AGN challenging to identify and characterize observationally. Such observational limitations explain why it has been so difficult to obtain statistically significant samples of dual AGN despite extensive observational efforts at, for example, optical/near-IR and X-ray wavelengths \citep{koss18, DeRosa2019,foord20}.

In the last few years, thanks to its outstanding sensitivity and spatial resolution, the Atacama Large Millimeter Array (ALMA) has generated a revolution in the study of molecular gas in major galaxy mergers. This is particularly true in the nearby Universe where high physical resolutions of $<$100\,pc are feasible. Examples of these studies include ALMA $^{12}$CO(1-0) observations of Arp\,220 \citep{scoville17}, a $^{12}$CO(2-1) mapping of the archetypical dual AGN NGC\,6240 at a $\sim$1 kpc projected nuclear separation \citep{treister20}, and more recently, the discovery of a dual AGN in UGC\,4211 at a record-breaking projected nuclear separation of $\sim$\,230 pc \citep{koss23}. These studies have consistently found evidence of large nuclear reservoirs of molecular gas with a wide variety of gas morphologies, as well as dynamical properties consistent with large-scale inflow and outflow motions.

Understanding the physical properties of these molecular gas reservoirs is critical to establish the evolutionary stage of the galaxy, as this is the fuel for both star formation activity and SMBH growth. Comparisons of the molecular gas influx rates and SMBH accretion rates of local major mergers, however, have found that the accretion of molecular gas onto the SMBH is a highly inefficient process \citep{muller-sanchez09, garcia-burillo14, treister18}. Specifically, the infalling gas must lose a significant fraction of its angular momentum prior to accretion, implying that the non-rotating low-velocity component of the molecular gas is the most likely contributor to SMBH growth. Studying the molecular gas kinematics within the SMBH sphere of influence (SoI) could, therefore, provide crucial insight into current and future SMBH growth, but this requires high angular resolution observations that can resolve the SoI. 

Here, we present high-resolution ($\sim$\,50$-$100 pc) ALMA observations of $^{12}$CO(2-1) or $^{12}$CO(1-0) emission in seven local galaxy mergers hosting dual AGN and single AGN. These low-$J$ $^{12}$CO transitions (rest-frame frequencies = 230.538 GHz and 115.271 GHz, respectively) are able to provide estimates of the total molecular gas mass and its distributions due to their low excitation temperatures, $\sim5$\,K, and low critical densities, $\sim10^3$\,cm$^{-3}$ (e.g., \citealt{papadopoulos12, bolatto13}). With these data, our main goals are to: (i) measure the mass of the molecular gas within the SoI, (ii) study the kinematic properties of the molecular gas within the SoI in relation to current and future SMBH growth, and (iii) understand the dynamics of the nuclear molecular gas. We will also briefly search for intrinsic differences between mergers hosting dual AGN and single AGN by comparing their molecular gas properties. This will help determine if the lack of simultaneous AGN activity in single AGN hosts is related to the supply of molecular gas available to the SMBH  (or if instead, their differences are explained by AGN variability (e.g., \citealp{sartori19}) and/or obscuration (e.g., \citealp{li21})). 

The paper is structured as follows: We present the sample in Section \ref{sec:sample}. We describe the observations as well as the data reduction, calibration, and imaging procedures are presented in Section \ref{sec:data}. We show the $^{12}$CO emission detections, the MCMC models of their kinematic features, and the derived molecular gas masses in Section \ref{sec:results}. In Section \ref{sec:disc}, we explore the correlations between the molecular gas properties and multi-wavelength AGN properties. We conclude our findings in Section \ref{sec:conclusion}. Throughout this paper, we assume a $\Lambda$CDM cosmology with $H_0 = 67$ km\,s$^{-1}$ Mpc$^{-1}$, $\Omega_M=0.315$, and $\Omega_\Lambda=0.685$ \citep{Planck2020}.

\section{Sample Selection}
\label{sec:sample}

The parent sample for this work comes from the Data Release 2 of the BAT Spectroscopic Survey (BASS; \citealt{Koss2022b}), composed of 858 AGN selected from the revised Swift-BAT 70-months all-sky catalog of ultra-hard, 14$-$195 keV, X-ray emitters. This ensures access to a variety of multi-wavelength measurements of AGN properties (e.g., black hole mass, AGN luminosities) that have been published by the BASS collaboration. From the BASS AGN sample, \citet{koss18} identified 30 local ($z$\,$\lesssim$\,0.05) major galaxy mergers, defined as those with mass ratios $<$\,5:1, with projected nuclear separations between 0.2 and 10\,kpc. Selecting from this set of major mergers, we restricted our final sample to systems with new or archival high resolution ($\sim 50-100$\,pc) ALMA detections of $^{12}$CO (2-1) and/or $^{12}$CO (1-0). 

Five dual AGN systems (NGC\,6240, Mrk\,463, Mrk\,739, UGC\,4211, and ESO\,253-G003) meet the above criteria, observed by ALMA during Cycles 8 through 10. This work focuses on the analysis of these objects. Seven single AGN systems also meet the aforementioned selection criteria, observed by ALMA during Cycles 8 and 10. In our sample, we include the two single AGN hosts (Mrk\,975 and NGC\,985) that were observed in Cycle\,8, but the remaining five Cycle\,10 objects will be analyzed in an upcoming work (Ramos et al., in prep.). Though not a complete comparison sample, the inclusion of these two single AGN will serve as a preliminary point of comparison to quantify any intrinsic differences between dual AGN and single AGN in merger systems. Below, we provide details of each system analyzed in this paper. A summary of the complete sample is also presented in Table~\ref{tab:sample}. Descriptions of the data and observations are presented in Section \ref{sec:data}.

\begin{deluxetable*}{lccccccc}
\tablenum{1}
\tablecaption{Properties of the Merging Galaxy Systems studied in this work\label{tab:sample}}
\tablewidth{0pt}
\tablehead{ \colhead{(1)} & \colhead{(2)} & \colhead{(3)} & \colhead{(4)} & \colhead{(5)} & \colhead{(6)}  \\
\colhead{System} & \colhead{Redshift} & \colhead{Distance} & \colhead{$\log$(L$_{\text{IR}}/$L$_\odot$)} & \colhead{Nuclear} & \colhead{Reference} \\[-0.2cm]
%& \multicolumn{2}{c}{Resolution}  \\
\colhead{Name} & \colhead{} & \colhead{(Mpc)} & \colhead{} & \colhead{Sep. (kpc)} & \colhead{}}
%\colhead{($\arcsec\times\arcsec$)} & \colhead{(kpc)}  }
%\decimalcolnumbers
\startdata
\multicolumn{6}{c}{Dual AGN}\\
\hline
NGC\,6240 & 0.0243 &110 & 11.9\tablenotemark{a} & 0.95 & \citet{komossa03} \\ %& 0.03 & 0.015 \\ 
Mrk\,463 & 0.0504 & 230 & 11.8\tablenotemark{b} & 3.8 & \citet{bianchi08} \\ %& 0.08 & 0.08\\
Mrk\,739 & 0.0299 & 	139 & 10.9\tablenotemark{d}& 3.4 & \citet{koss11} \\ %& 0.08 & 0.04 \\
UGC\,4211 & 0.0344 & 150 & 10.7\tablenotemark{c}& 0.23 & \citet{koss23} \\ %& 0.06 & 0.04  \\
%LEDA 17883 & 0.0503 & 224 & & 0.91  && 2022.1.01348.S \\ %& 0.05 & 0.051\\
ESO\,253-G003 & 0.0425 & 188 & 11.4\tablenotemark{e} & 1.4 & \citet{Tucker2021}\\ %& 0.05 & 0.044\\
\hline\hline
\multicolumn{6}{c}{Single AGN}\\
\hline
Mrk\,975 & 0.0496 & 215 &  11.5\tablenotemark{e} & 2.5 & \citet{koss18} \\ %&  0.09 &  0.095\\
NGC\,985 & 0.0431 & 190 & 11.3\tablenotemark{f}& 3.3 & \citet{appleton93} \\ %& 0.13 & 0.117
\enddata
\tablecomments{Column (1): Name of system. Column (2): Redshift. Column (3): Distance to source in megaparsecs. Column (4): Logarithm of total infrared luminosity in solar luminosity units. (5): Projected nuclear separation of SMBHs in kiloparsecs. Column (6): Reference for projected nuclear separation. $^a$\citet{howell10}; $^b$\citet{imanishi14}; $^c$\citet{shimizu17}; $^d$\citet{koss11}; $^e$\citet{bonatto97}; $^f$\citet{appleton02}}
%\tablenotetext{a}{\citet{howell10}}
%\tablenotetext{b}{ \citet{imanishi14}}
%\tablenotetext{c}{ \citet{shimizu17}}
%\tablenotetext{d}{ \citet{koss11}} 
%\tablenotetext{e}{ \citet{bonatto97}}
%\tablenotetext{f}{ \citet{appleton02}}
\end{deluxetable*}

\subsection{NGC\,6240}

The Luminous Infrared Galaxy (LIRG) NGC\,6240 is a late-stage merger galaxy that is often considered the prototypical dual AGN system. \textit{Chandra}  X-ray observations have confirmed the presence of two AGN in NGC 6240, separated by 1.7$''$ or 1.4\,kpc \citep{komossa03, wang14}. Column density estimates from \textit{NuSTAR} observations find that both nuclei are Compton Thick ($N_H$\,$>$\,$10^{24}$\,cm$^{-2}$; \citealp{Puccetti2016}), indicating that the two SMBHs are likely rapidly growing in this phase of heavy nuclear obscuration. This is a highly turbulent system with large amounts of molecular gas residing between the two nuclei \citep{tacconi99, engel10, feruglio13, treister20} and strong feedback mechanisms driven by both starbursts and AGN \citep{muller-sanchez18, Cicone2018,Ceci2025}.

\subsection{Mrk\,463}

Mrk\,463 is a nearby ($d$\,$\simeq$\,230\,Mpc) Ultra-Luminous Infrared Galaxy (ULIRG). Confirmed as a dual AGN by {\it Chandra} X-ray observations 
\citep{bianchi08}, the two nuclei are separated by 3.8$\,\pm$\,0.01$''$, or $\sim$\,3.8 kpc. The SMBH in the Mrk\,463E nucleus is undergoing rapid growth and has not reached the deeply buried stage ($N_H = 8 \times 10^{23}$\,cm$^{-2}$; \citealt{yamada18}), while Mrk\,463W is at an early stage of SMBH growth ($N_H = 3 \times 10^{23}$\,cm$^{-2}$; \citealt{yamada18}). Optical \textit{Multi-Unit Spectroscopic Explorer} (MUSE) Integral Field Unit (IFU) observations of Mrk\,463 have also revealed the presence of tidal tails, star-forming clumps, and emission line regions \citep{treister18}. Mrk\,463 likely represents an earlier stage of a major galaxy merger, eventually leading to the heavily obscured/coalescence phase of sources like NGC\,6240 and Arp\,220. 

\subsection{UGC\,4211}

At a projected nuclear separation of only 0.31$''$, or 230 pc at 150 Mpc, UGC\,4211 is the smallest separation system in the sample of {\it Swift}/BAT hard X-ray selected AGN from \citet{koss18}. It has total  X-ray-derived obscuration of $N_H$\,=\,9\,$\times$\,10$^{22}$\,cm$^{-2}$ \citep{zhao21}. UGC\,4211 was recently confirmed as a dual AGN by {\it HST}/STIS data and ALMA millimeter-wave continuum observations \citep{koss23}. In addition to the ALMA observations presented in this work, this system has archival Keck/AO imaging and near-IR IFU spectra, Jansky Very Large Array 22 GHz continuum images, {\it Chandra} X-ray images, and \textit{MUSE} NFM IFU data available \citep{koss18, koss23}. The $^{12}$CO(2-1) transition was previously observed at a lower angular resolution by Institut de Radioastronomie Millim\'etrique (IRAM; \citealt{koss21}; Shimizu et al., in prep).

%The two SMBHs have similar masses of log($M_{BH}/M_\odot$) $\sim$ 8.1 (south) and log($M_{BH}/M_\odot$) $\sim$ 8.3, respectively \citep{koss23}.

\subsection{Mrk 739}

Mrk\,739 is a nearby ($z$\,=\,0.0299) late-stage merger galaxy that hosts two nuclei (Mrk\,739E and Mrk\,739W) separated by 3.4 kpc \citep{koss11}. Mrk\,739E was identified as a type I AGN due to the broad emission line components in its optical spectrum \citep{Netzer1987, koss11, Tubin2021}. While the optical spectrum of Mrk\,739W contains only narrow emission lines \citep{Tubin2021}, \textit{Chandra} observations by \citet{koss11} identified an AGN in Mrk\,739W based on its hard photon index in the soft X-ray (0.5$-$10\,keV) band. While many late-stage mergers host buried AGN, Mrk\,739E is unabsorbed ($N_H < 6.5 \times 10^{19}$\,cm$^{-2}$; \citealt{Inaba2022}), and Mrk\,739W is only weakly absorbed ($N_H \simeq 6.9 \times 10^{21}$\,cm$^{-2}$; \citealt{Inaba2022}). 

%The two nuclei are moderate-to-low luminosity with intrinsic 2$-$10 keV X-ray luminosities of $1.0 \times 10^{43}$\,erg\,s$^{-1}$ and $7.5 \times 10^{41}$\,erg\,s$^{-1}$ for Mrk\,739E and Mrk\,739W, respectively \citep{Inaba2022}. If the system is in a late merger stage, these unique AGN properties counter current theories of black hole fueling, making Mrk\,739 an interesting galaxy to include in our sample. 

\subsection{ESO\,253-G003}

The LIRG ESO\,253-G003 ($z$\,=\,0.0425) is a late-stage merger with kpc-scale tidal arms. The galaxy hosts two AGN separated by 1.4 kpc (1.7$''$) as identified by recent \textit{MUSE} IFU spectra \citep{Tucker2021}. The brighter northeastern nucleus is a known AGN \citep{vc2010} with asymmetric broad emission-line profiles \citep{Tucker2021}. Periodic outbursts have been observed from this brighter AGN, likely caused by a repeated tidal disruption event \citep{Payne2021}. In the fainter southwestern nucleus, the spectrum of shows emission line ratios and high ionization line detections consistent with a second AGN \citep{Tucker2021}, thus confirming the system as a dual AGN. In addition to its disturbed morphology, MUSE IFU observations of ESO\,253-G003 showed indicators of kpc-scale outflows and a possible AGN-driven superbubble \citep{Tucker2021}. 

\subsection{Single AGN in late-stage mergers}

As a tentative point of comparison to the dual AGN systems, we include two single AGN in our sample, NGC\,985 and Mrk\,975. These systems are in similar evolutionary stages of the galaxy merger as the dual AGN systems presented above. Mrk\,975 is 215\,Mpc away ($z$\,=\,0.0496) and has a projected nuclear separation of 2.5 kpc (2.5$''$), while NGC 985 is at a distance of 190 Mpc ($z$\,=\,0.0431) with a projected nuclear separation of 3.3 kpc (3.7$''$). Hereafter, we refer to the inactive SMBHs in these single AGN systems as non-AGN SMBHs. This is to emphasize that while AGN activity has not been detected, these non-detections could be an obscuration effect and/or the SMBHs could be accreting at a low rate.

\section{Data Description}
\label{sec:data}

The data analyzed in this work come from three ALMA programs targeting $^{12}$CO emission in AGN-hosting major mergers at $\sim50-100$ pc resolutions, which we describe here. Generally, for systems that have been observed during multiple ALMA observing cycles, we analyzed the most recent data set. For only ESO\,253-G003, both $^{12}$CO(1-0) and $^{12}$CO(2-1) transitions had been observed, but we prioritized the $^{12}$CO(1-0) transition for which the line luminosity can be directly converted into a molecular gas mass, whereas the $^{12}$CO(2-1) transition would require assuming a line ratio (see Section \ref{subsec:CO_prop}). In this section, we report the observational setup, data reduction, and the imaging procedures performed on these data.  The image properties of the resulting ALMA $^{12}$CO data cubes are presented in Table \ref{tab:obs}.

\subsection{ALMA Cycle\,8 and Cycle\,10 Observations: $^{12}$CO(2-1)}
\label{subsec:cycle8-10}

The $^{12}$CO(2-1) observations presented in this work were obtained as part of Programs 2021.1.01019.S and 2023.1.01196.S (PI: E. Treister) from ALMA Cycles 8 and 10, respectively. These programs obtained 11 and 24.2 hours of ALMA Band 6 observing time, respectively. The observations targeted a combined total of eleven major galaxy mergers at high angular resolutions (0.05$-$0.1$"$). Of these systems, four are confirmed dual AGN (Mrk\,463, Mrk\,739, UGC\,4211, and ESO\,253-G003).  We analyze these $^{12}$CO(2-1) data in this work, except for ESO\,253-G003 for which we have Cycle\,9 observations of $^{12}$CO(1-0) that allow for a more direct estimate of the molecular gas mass (discussed in Section \ref{subsec:cycle9}). We also include the two single AGN, NGC\,985 and Mrk\,975, that were observed in the Cycle\,8 program. 

Data were obtained using two different array configurations for all targets: a high-resolution antenna setup (TM1) and an intermediate resolution antenna set-up (TM2). Combining the observations in these two configurations allows us to maintain high angular resolutions ($\sim$\,0.05$-0.1''$ ), while achieving a maximum recoverable scale of $\sim$3$-5''$. Observational details for individual targets (e.g., the antenna configurations, observation dates, and on-source observing times) are presented in Appendix \ref{app:obs_details}.

% In all cases, we defined a total of four spectral windows. The two main windows were contiguous and overlapping, centered on the expected position of the $^{12}$CO(2-1) at the redshift of the system and thus covered 

In all cases, the spectral setup has four spws (1.875 GHz wide) with the $^{12}$CO(2-1) emission covered by two contiguous and overlapping spectral windows. This setup allow us to cover velocity offsets up to $\pm$\,2000 km s$^{-1}$ from the spectral line. An additional spectral window was centered on the CS(5-4) line. Though detections of CS(5-4) were not expected because of its typical faintness, this line traces dense molecular gas with a critical density approximately three orders of magnitude higher than that of the $^{12}$CO(2–1) line. The final spectral window was placed around $\sim$247 GHz in the observed frame to trace the continuum emission, free of any emission lines. The spectral windows containing the $^{12}$CO(2-1) line were configured to use $\sim$\,2 MHz ($\sim$\,3 km s$^{-1}$) channel widths. For the CS(5-4) window and continuum window, 7.8 MHz ($\sim$\,10 km s$^{-1}$) and 31 MHz ($\sim$\,40 km s$^{-1}$) channel widths were used in Cycle\,8. In Cycle\,10, 41.67 MHz ($\sim$\,53 km s$^{-1}$) channel widths were used. 

Data reduction was carried out for each target using the ALMA pipeline version 2021.2.0.128 (CASA version 6.2.1.7) and 2023.1.0.124 (CASA version 6.5.4.9) for Cycles 8 and 10, respectively \citep{CASA2022, ALMA2023}. We combined the TM1 and TM2 observations with the \texttt{concat} tool in CASA version 6.5.1. We then fit the continuum emission from the spectral line-free channels across the four spectral windows using CASA task \texttt{uvcontsub}, assuming a zeroth-order polynomial. Finally, we generated a continuum-subtracted $^{12}$CO(2-1) emission line data cube using \texttt{tclean}. Each data cube was imaged with Briggs weighting (robust=0.5; \citealt{briggs95}). To improve the signal-to-noise of each channel, we smoothed the channel widths to 7.8 MHz ($\sim$\,10 km s$^{-1}$) in the final data cube. The resulting beam sizes and 1$\sigma$ RMS values of each target are reported in Table \ref{tab:obs}. 

\begin{deluxetable}{ccccccc}
\tablenum{2}
\tablecaption{ALMA $^{12}$CO Data Cube Properties\label{tab:obs}}
\tablewidth{0pt}
\tablehead{ \colhead{(1)} & \colhead{(2)} & \colhead{(3)} & \colhead{(4)}  & \colhead{(5)}  & \colhead{(6)} & \colhead{(7)}\\[-0.2cm]  
\colhead{System} & \colhead{ALMA Band} & \colhead{$^{12}$CO transition} & \colhead{Beam Size} & \colhead{Beam Size} & \colhead{1$\sigma$ RMS}   & \colhead{Program ID} \\[-0.2cm]
\colhead{} & \colhead{} & \colhead{} & \colhead{($"\times ", ^{\circ}$)} & \colhead{(pc $\times$ pc)} & \colhead{(mJy/beam)}    & \colhead{}} 

%\decimalcolnumbers
\startdata
\multicolumn{6}{c}{Dual AGN}\\
\hline
NGC\,6240 & Band 6 & $^{12}$CO(2-1) & 0.06$\times$0.03, $-72^{\circ}$ & 31$\times$15 & 0.28 & 2015.1.00370.S \\
Mrk\,463 & Band 6 & $^{12}$CO(2-1)& 0.13$\times$0.11, 8.6$^{\circ}$ & 122$\times$112 & 0.38 & 2023.1.01196.S \\
Mrk\,739 & Band 6 & $^{12}$CO(2-1) & 0.12$\times$0.10, 16$^{\circ}$ & 80$\times$68 & 0.33 & 2023.1.01196.S \\
UGC\,4211 & Band 6 & $^{12}$CO(2-1) & 0.07$\times$0.06, 0.6$^{\circ}$ & 53$\times$44 &  0.30 & 2021.1.01019.S \\
%LEDA 17883 & 0.09$\times$0.05, 57$^{\circ}$ & 94$\times$56 & 0.24\\
ESO\,253-G003 & Band 3 & $^{12}$CO(1-0) & 0.06$\times$0.05, 70$^{\circ}$ & 52$\times$44 & 0.17  & 2022.1.01348.S \\ 
\hline\hline
\multicolumn{6}{c}{Single AGN}\\
\hline
Mrk\,975 & Band 6 & $^{12}$CO(2-1) & 0.10$\times$ 0.09, 35$^{\circ}$ & 107$\times$90 & 0.16  & 2021.1.01019.S  \\
NGC\,985 & Band 6 & $^{12}$CO(2-1) & 0.14$\times$ 0.12$, -79^{\circ}$ & 124$\times$114 & 0.71 & 2021.1.01019.S 
\enddata
\tablecomments{Column (1): Source name. Column (2): ALMA continuum band. Column (3): Observed $^{12}$CO transition.  Column (4): Synthesized beam size in arcseconds with position angle in degrees. Column (5): Projected synthesized beam size in parsecs. Column (6): RMS sensitivity of data cubes, defined as mean 1$\sigma$ noise level per 10 km/s channel. Column (7) ALMA Program ID for data analyzed in this work. }
\end{deluxetable}

\subsection{ALMA Cycle\,9 Observations: $^{12}$CO(1-0)}
\label{subsec:cycle9}

Twelve local post-merger galaxies were observed by ALMA as a part of Program 2022.1.01348.S (PI: E. Treister), for a total of 35.4 hours at Band 3. These observations targeted the $^{12}$CO(1-0) at angular resolutions of $\sim0.05"$ (maximum recoverable scales of $\sim0.5"$). Of the observed targets, we detect $^{12}$CO(1-0) emission in only two galaxies: LEDA\,17883 and the dual AGN system ESO\,253-G003. While we include ESO\,253-G003 in our dual AGN sample, we omit LEDA 17883 from this study because its single or dual AGN nature remains highly debated (del Moral-Castro et al., in prep.). 

The data were obtained using the C43-9 antenna set-up (TM1) with the longest baseline configuration of $\sim$15 km. The exact observing dates and on-source observing times of each execution block are provided in Appendix \ref{app:obs_details}. The receiver's setup consisted of four spectral windows, with three targeting the 100\,GHz mm-wave continuum emission. Though the observing sensitivities were set based on the expected continuum brightness at 100\,GHz, a fourth spectral window was centered on the redshifted frequency of the $^{12}$CO(1-0) spectral line. All spectral windows were configured to have a total bandwidth of 1.875 GHz ($\sim$5000 km s$^{-1}$) with spectral resolutions of 1.129 MHz ($\sim$3 km s$^{-1}$). 

We performed the data reduction and calibration using the ALMA pipeline version 2022.2.0.64 (CASA version 6.4.1.12) for all targets \citep{CASA2022, ALMA2023}. Continuum fitting and subtraction were performed using \texttt{uvcontsub}. The \texttt{tclean} task was then used to produce a continuum-subtracted cube containing only emission from the $^{12}$CO(1-0) line. Since observations were only conducted in the compact array configuration (TM1) for these targets, imaging was performed using natural weighting (robust = 2) to maximize sensitivity to diffuse emission. The channel widths were smoothed to 7.8 MHz ($\sim$20 km s$^{-1}$) for analysis purposes. 

\subsection{Existing NGC 6240 ALMA Observations }

We included archival high-resolution ALMA Band 6 observations of the archetypical dual AGN NGC\,6240 from Cycle\,3 (Program 2015.1.00370.S; PI: E. Treister). These observations targeted the $^{12}$CO(2-1) line and was first presented by \citet{treister20}. Briefly, the spectral configuration used in this program was similar to those of our Cycle\,8 and Cycle\,10 programs, covering the $^{12}$CO(2-1) and CS(5-4) lines and the surrounding continuum. The targeted spatial resolution is slightly better ($\sim$0.03$''$) since baselines up to $\sim$15 km were used. However, the resulting maps and data cubes have spatial resolutions and noise levels that are comparable to our Cycle\,8 data. These properties are reported in Table \ref{tab:obs}. 

\subsection{Supporting Multi-wavelength Information}

The systems in our sample have been the subject of extensive multi-wavelength observations in the past, allowing us to compile useful information that supports our work. In Table \ref{tab:phys_prop}, we present a compilation of the physical properties of the sources in our sample.

%It is important, however, to note that due to their small projected nuclear separations and their associated complexity, some of the quantities reported here are integrated values and do not resolve the two components of the system. 

\subsubsection{Stellar Masses}
For most of our targets, global stellar masses were reported by \citet{koss21}. They were derived by combining 2MASS photometry in the near-IR, which is more sensitive to stellar emission in merger systems, with mid-IR data obtained from the AllWISE catalogs \citep{wright10}, which is more sensitive to AGN emission. A detailed description of the procedure followed to estimate stellar masses can be found in \citet{powell18a}. These measurements correspond to the system's total stellar mass, which can then be combined with the ratios of the masses for each component, as measured and reported by \citet{koss18}. Similarly, for ESO\,253-G003, a global stellar mass measurement is reported in \citet{Payne2021}, which was determined by SED (spectral energy distribution) fitting. The SED consisted of fluxes from the 2MASS All-Sky Point Source Catalog in the near-IR \citep{Skrutskie2006}, the AllWISE Source Catalog in the mid-IR \citep{wright10}, and the IRAS Faint Source Catalog in the far-IR \citep{Moshir1990}. A mass ratio, however, is not available in the current literature, so we can only report the total stellar mass of ESO\,253-G003. Finally, \citet{Tubin2021} presented stellar mass estimates derived from the AGN-corrected K-band magnitudes of each component of Mrk\,739. These measurements are summarized in Table \ref{tab:phys_prop}.

\subsubsection{Black Hole Masses \& Sphere of Influences}
Masses of the central SMBHs  were obtained from the literature where available. This is the case for NGC\,6240, UGC\,4211, and Mrk\,739. Specifically, \cite{medling15} reports SMBH masses $M_{\text{BH}}$ for NGC\,6240, determined by 2D stellar kinematic maps and models of the mass density profile.  For Mrk\,739, \citet{Tubin2021} derived the $M_{\text{BH}}$ of each nuclei using the $H\beta$ line width and luminosity (procedure outlined in \citealt{vestergaard06}). Finally, \citet{koss23} calculated the $M_{\text{BH}}$'s of UGC 4211 from their stellar velocity dispersions $\sigma_*$ by applying the $M_{\text{BH}}-\sigma_*$ relation,
\begin{equation}
\label{eq:MBH_disp}
\text{log}\left(M_{\text{BH}}/M_\odot\right) = 4.38 \times \text{log}\left(\frac{\sigma_*}{200 \text{ km s$^{-1}$}}\right)+8.49,
\end{equation}
\noindent from \citet{kormendy13}. For the remaining nuclei (Mrk\,463, ESO\,253-G003, NGC\,985, and Mrk\,975), we follow the procedure of \citet{koss23} and employ the $M_{\text{BH}}-\sigma_*$ relation above to calculate the $M_{\text{BH}}$. For Mrk\,463E, $\sigma_*$ was determined by \citet{dasyra11}. For Mrk\,463W and both nuclei of ESO\,253- G003, we use archival VLT/MUSE IFU data obtained in the Narrow Field Mode (ESO Program 0104.B-0497; PI: E. Treister) and the Wide Field Mode (ESO Program 096.D-0296; PI: J. Anderson), respectively. These MUSE data were automatically reduced using the instrument pipeline \citep{weilbacher20} and had angular resolutions of $\sim$0.1$''$ and $\sim$0.7$''$, respectively. These angular resolutions were sufficiently high to isolate the nuclear regions of the merging systems. We then measured $\sigma_*$ using the Ca II triplet absorption lines in the rest-frame 8350-8730\AA~range, following the method outlined by \citet{koss22}.

Similarly, individual SMBH masses for the single AGN systems could be estimated based on existing VLT/MUSE IFU data. Both NGC 985 and Mrk 975 were observed in the Wide Field Mode as a part of ESO Programs 094.B-0345 (PI: B. Husemann; \citealp{husemann22}) and 114.27GQ.001 (PI: E. Treister), respectively. Following the same procedure as Mrk\,463 and ESO\,253-G003 described above, our team obtained and analyzed the pipeline-reduced MUSE data cubes to obtain the SMBH mass values.

We note as an important caveat that SMBH mass measurements can be highly uncertain, and vary substantially based on the method followed (e.g., \citealt{Gebhardt2011,Thater2022}). For example, \citet{Thater2020} found that ionized and molecular gas-based measurements of $M_{\text{BH}}$ are systematically lower than those derived from the stellar dynamics (see their Figure 2). These discrepancies can be attributed to the fact that each method is built upon a different set of assumptions and probes the gravitational potential of the SMBH in different ways. We aim to minimize the systematic biases in our $M_{\text{BH}}$ measurements by utilizing the same method when possible. Specifically, by utilizing Equation \ref{eq:MBH_disp} in accordance with \citet{koss23},  we ensure that the majority ($\sim$70\%) of the SMBH masses used in this work were measured via the same method. 

With these SMBH masses, we can derive a crucial quantity of interest: the SMBH SoI. If we assume that the nuclei have virialized, isotropic, and spherically symmetric gas motions (e.g., \citealt{VDBosch16}), then we can estimate the radius of the SoI $r_{\text{SoI}}$ using the definition $r_{\text{SoI}} = GM_{\text{BH}}\sigma_*^{-2}$ \citep{koss22}. Following the procedure of \citet{koss22}, we apply the $M_{\text{BH}}-\sigma_*$ relation (Equation \ref{eq:MBH_disp}) to obtain a simplified prescription,

\begin{equation}
r_{\text{SoI}}=33 \textrm{pc} \left(\frac{\sigma_*}{200 \textrm{ km s}^{-1}}\right)^{2.38}. 
%\approx 0.0008 \text{pc} \left( \frac{M_{\text{BH}}}{M_\odot}\right)^{0.543}.
\label{eq:SOI}
\end{equation}

We note, however, that there is evidence that the properties of these complex galaxy mergers do not align with the established $M_{\text{BH}}-\sigma_*$ relation. In general, gas-rich mergers tend to have higher $M_{\text{BH}}$ than what is inferred from the relation \citep{medling15}, with divergences as high as a factor of 2.5 \citep{Ruby2024}.  Thus, Equation \ref{eq:SOI} could be underestimating the SoI of the SMBHs in our sample. Further uncertainty is introduced by the mixed definitions of the SMBH SoI that are offered in the current literature. While Equation \ref{eq:SOI} defines the SoI as the region in which the SMBH gravity dominates over that of the host, assuming a virialized system that also follows the $M_{\text{BH}}-\sigma_*$ relation, other studies define it more generally as the region in which the SMBH contributes over 50\% of the mass (e.g., \citealt{medling15}). These discrepancies can result in significant and systematic differences between SoI measurements. For example, \citet{medling15} conducted Markov Chain Monte Carlo (MCMC) modeling of the mass profiles of both nuclei of NGC 6240 and found substantially greater SoIs ($r_{\text{SoI}} \sim250$ pc) than the radii inferred from Equation \ref{eq:SOI} ($\sim60$ pc). They similarly report SoI radii that are larger than the $M_{\text{BH}}-\sigma_*$ relation value by factors of $\sim1-4$ for other gas-rich major mergers hosting AGN \citep{medling15}. 

We therefore estimate that there can be up to a factor of $\sim4$ difference in the derived SoI based on methodology and definition. To account for this uncertainty, we define upper- and lower-bound estimates of $r_{\text{SoI}}$ and conduct the same analysis at both extrema. Here, we use Equation \ref{eq:SOI} as the lower-bound of $r_{\text{SoI}}$. Based on the \citet{medling15} results for NGC\,6240, we apply a factor of $\sim4$ to the lower-bound ($M_{\text{BH}}-\sigma_*$ SoI) to obtain our upper-bound SoI estimate.

\subsubsection{Bolometric Luminosities}
Another relevant quantity for this study is the AGN bolometric luminosity of each individual nucleus. For Mrk 739 and UGC\,4211, AGN bolometric luminosities of each nucleus were reported in \citet{koss11} and \citet{koss23}. For NGC\,6240, Mrk\,463, NGC\,985 and Mrk\,975, the intrinsic 2$-$10 keV X-ray luminosities of the individual nuclei were available in the literature \citep{Ricci2017c, Ricci2021, bianchi08} which we convert into an AGN bolometric luminosity by assuming a bolometric correction of $\kappa_{2-10} = 20$ for the $2-10$ keV bands ($L_{\text{Bol}}$ = $\kappa_{2-10} \times$ $L_{2-10}$; \citealt{Vasudevan2009, Gupta2024}). For the dual AGN system ESO\,253-G003, we estimate intrinsic 2$-$10 keV X-ray luminosities from the ALMA continuum luminosities at $\sim$100\,GHz (Droguett et al., in prep). At our $<$100\,pc resolution, self-absorbed synchrotron emission from the AGN corona dominates the $\sim$100\,GHz continuum, given that the contribution of diffuse star formation emission is minimized.  This allows for a robust direct conversion into an X-ray luminosity as established by the millimeter-X-ray relation for AGN \citep{Kawamuro2022, Ricci2023}. We once again assume a bolometric correction of $\kappa_{2-10} = 20$ \citep{Vasudevan2009, Ricci2023} to obtain estimates of the AGN bolometric luminosities for each nucleus. While other studies have reported a luminosity dependence of the bolometric correction \citep[e.g.,][]{Duras2020}, in the luminosity range covered by our sample, $\sim$5$\times$10$^9-$4$\times$10$^{11}$$L_\odot$, we obtain similar and consistent values for the bolometric correction.

\subsubsection{Eddington Ratios}
We derive Eddington ratios ($L_{\text{Bol}}/L_{\text{Edd}}$) for the individual nuclei using the aforementioned AGN bolometric luminosities and SMBH masses, defining the Eddington luminosity as:  
\begin{equation}
    L_{\text{Edd}} = 1.5 \times 10^{46} \text{ erg s}^{-1} \frac{M_{\text{BH}}}{10^8 M_\odot}.
\end{equation}
\noindent This definition is in accordance with \citet{Koss2022b} and assumes solar metallicity. 

\begin{deluxetable*}{lcccccccc}
\tablenum{3}
\tablecaption{Physical properties of the systems studied in this work \label{tab:phys_prop}}
\tablewidth{0pt}
\tablehead{ \colhead{(1)} & \colhead{(2)} & \colhead{(3)} &  \colhead{(4)} &  \colhead{(5)} & \colhead{(6)} & \colhead{(7)}   & \colhead{(8)} \\
\colhead{System} & \colhead{Stellar Mass} & \colhead{Mass} &  \colhead{SMBH Mass} &  \colhead{SoI LL} & \colhead{SoI UL} & \colhead{AGN Lum.}   & \colhead{Eddington} \\[-0.2cm]
\colhead{} & \colhead{$\log$($M_*/M_\odot$)} & \colhead{Ratio} &\colhead{$\log$($M_*/M_\odot$)} &  \colhead{Radius (pc)} & \colhead{Radius (pc)} & \colhead{log($L_{\text{bol}}$/erg s$^{-1}$)}  & \colhead{Ratio}
}
%\decimalcolnumbers
\startdata
\multicolumn{8}{c}{Dual AGN}\\
\hline
NGC\,6240 &  11.328 $\pm$ 0.0933\tablenotemark{a} &  2.09\tablenotemark{c} \\ %& 229 $\pm$ 43 & 9.2 & 45.54 & 44.62  & 0.0046 \\ %4.17 $\times 10^{44}$
North & 11.02 & -- & 8.94 $\pm$ 0.02\tablenotemark{d} & 58.0 & 235\tablenotemark{d} & 44.60\tablenotemark{e} & 0.0030 \\ %&  254 $\pm$ 3
South & 10.71 & -- &  8.94$^{+0.04}_{-0.005}$\tablenotemark{d} & 58.0 & 212\tablenotemark{d} & 45.02\tablenotemark{e} & 0.0080\\ %253$^{+5}_{-0.7}$ stellar visp
Mrk\,463  & 10.895 $\pm$ 0.0896\tablenotemark{a} & 3.63\tablenotemark{c} & \\ %& 6.72 & & 5.62 $\times 10^{44}$ & 0.7079 \\
East & 10.03 & --  & 7.74\tablenotemark{f} & 12.9 & 52.4 & 44.48\tablenotemark{g} & 0.0364 \\ %& 134.83
West & 10.59 & -- & 7.93 & 16.4 & 66.4 & 43.88\tablenotemark{g} & 0.0060 \\ %& 149
Mrk\,739 & 11.02 & 2.2\tablenotemark{c} & & & & \\ %1.82 $\times 10^{44}$ & 0.1259 \\
East & 10.86\tablenotemark{h} & --  & 7.22 $\pm$ 0.25\tablenotemark{h} & 6.7 & 27.3 & 45.0\tablenotemark{h} & 0.4017 \\ %&  93
West &  10.50\tablenotemark{h} & --  &  6.68 $\pm$ 0.045\tablenotemark{h} & 3.4 & 13.9 & 43.3\tablenotemark{h} & 0.0220 \\ %&  91 $\pm$ 5 
UGC\,4211 & 11.072 $\pm$ 0.0827\tablenotemark{a}  &  3.63\tablenotemark{c} & & \\ %8.24 & & 3.31 $\times 10^{44}$ & 0.0126 \\
North & 10.77 & -- & 8.4\tablenotemark{i} & 29.5 & 119.5 & 43.8 $\pm$ 1.8\tablenotemark{i} & 0.0018 \\ %& 200 $\pm$ 14
South & 10.21 & -- & 8.1\tablenotemark{i} & 20.3 & 82.2 & 45.3 $\pm$ 1.8\tablenotemark{i} & 0.1057 \\ %& 165 $\pm$ 17
%LEDA 17883 & 11.03 & & & 8.19$\pm$0.16  & 22.73 & 3.55 $\times 10^{44}$ & 0.0155\\
%East &  &  &  & & \\
%West &  &  &  & & \\
ESO\,253-G003 & 10.566\tablenotemark{b} & -- & & & & \\ %5.12 $\times 10^{44}$ & 0.0537 \\
NE &  -- & -- & 7.3 & 8.3 & 33.7 & 45.16\tablenotemark{j} & 0.4776 \\ %& 127
SW &  -- & -- & 7.9 & 17.4 & 70.7 & 45.05\tablenotemark{j} & 0.0948 \\ %& 153
\hline\hline
\multicolumn{8}{c}{Single AGN}\\
\hline
Mrk\,975 & 10.98 $\pm$ 0.083\tablenotemark{a} &  2.29\tablenotemark{c} & & & & \\ %6.79$\times$10$^{44}$  & 0.0669\\
NE & 10.32 & -- & 7.0 & 5.1 & 20.7 & 44.86\tablenotemark{k} & 0.4830 \\ %& 91
NW$^{*}$ & 10.68 & -- & 7.9 & 15.8 & 64.0 & -- & --  \\ %& 141
NGC\,985 & 11.11 $\pm$ 0.082\tablenotemark{a} & 6.9\tablenotemark{c} & & & &  \\ %9.29$\times$10$^{44}$ & 0.0481\\
East & 10.80 & -- & 9.5 & 116.8 & 473.5 & 45.08\tablenotemark{k} & 0.0025 \\ %& 340
West$^{*}$ & 9.96 & -- & 8.0 & 17.9 & 72.5 & -- & -- \\ %& 155 
\enddata
\tablenotetext{*}{non-AGN}
\tablecomments{Column 1: Name of system and its components. Column 2: Logarithm of the total stellar mass. Column 3: Mass ratio of the components. This value was not available in the literature for ESO\,253-G003. Column 4: Logarithm of the SMBH mass. Column 5: Lower-bound estimate of the SMBH sphere of influence. Column 6: Upper-bound estimate of the SMBH sphere of influence. Column 7: Logarithm of the AGN bolometric luminosity in erg s$^{-1}$. Column 8: Eddington ratio. References: $^a$\citet{koss21}; $^c$\citet{koss18}; $^b$\citet{Payne2021}; $^d$\citet{medling15}; $^e$\citet{Ricci2021}; $^f$\citet{dasyra11}; $^g$\citet{bianchi08}; $^h$\citet{Tubin2021}; $^i$\citet{koss23}; $^j$Droguett et al. (in prep); $^k$\citet{Ricci2017c};}
\end{deluxetable*}

\section{Results}
\label{sec:results}

\subsection{CO Maps and Properties} \label{subsec:CO_prop}

\begin{figure*}
    \centering
    \includegraphics[width=\textwidth]{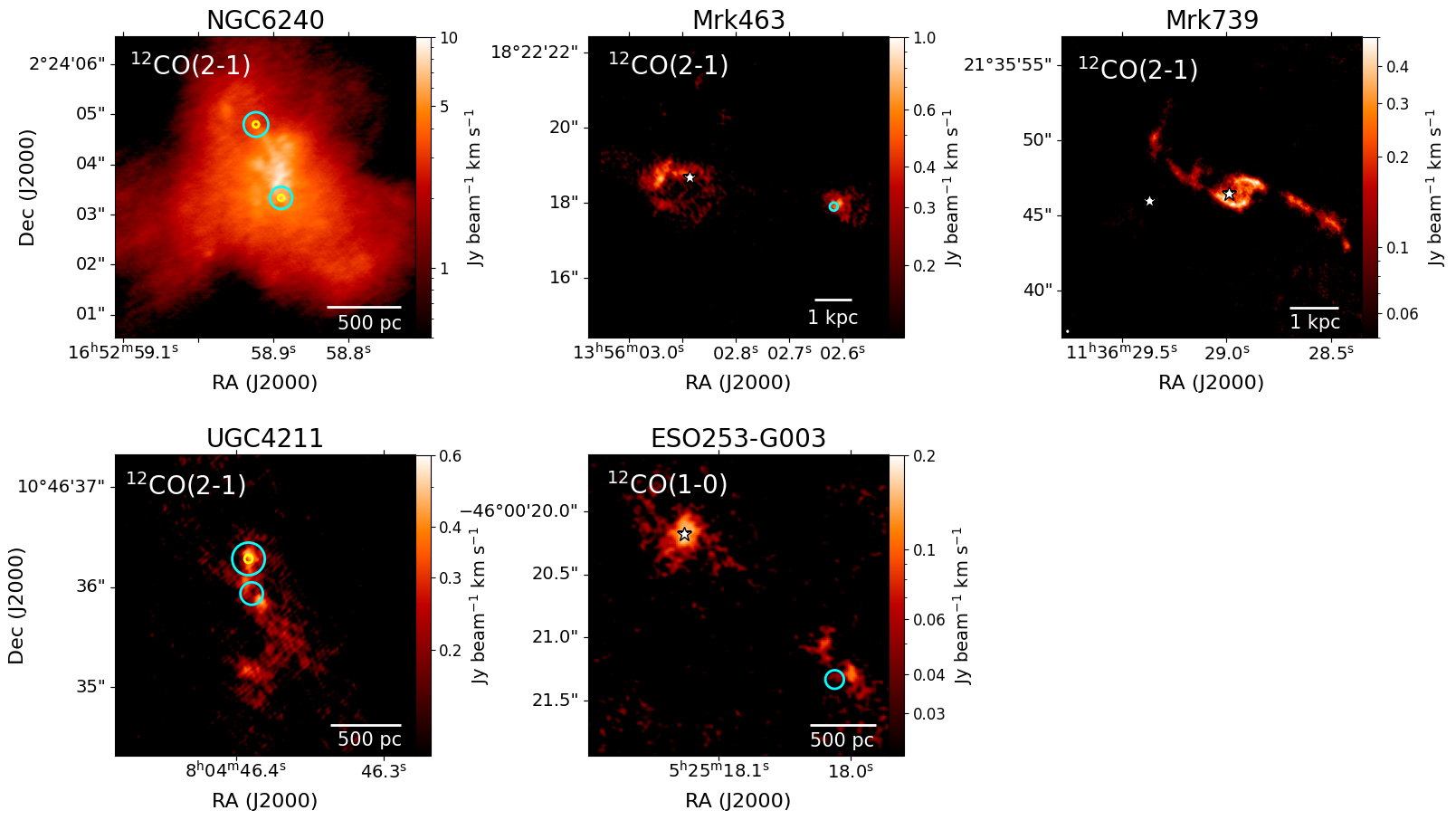}
    \caption{High-resolution $^{12}$CO moment 0 (integrated intensity) maps of confirmed dual AGN in major mergers. When resolved, upper-bound and lower-bound estimates of SMBH SoI are overlaid as cyan and yellow circles, respectively. If only the SoI upper-bound is resolved, we only show the upper-bound. For systems with unresolved SoIs, we mark the AGN position with a white star. Here, we analyze $^{12}$CO(1-0) for ESO\,253-G003 emission, whereas the remaining systems only have $^{12}$CO(2-1) detections. Note that the colormap is shown on a logarithmic scale.}
    \label{fig:mom0_double}
\end{figure*}

\begin{figure*}
    \centering
    \includegraphics[width=\textwidth]{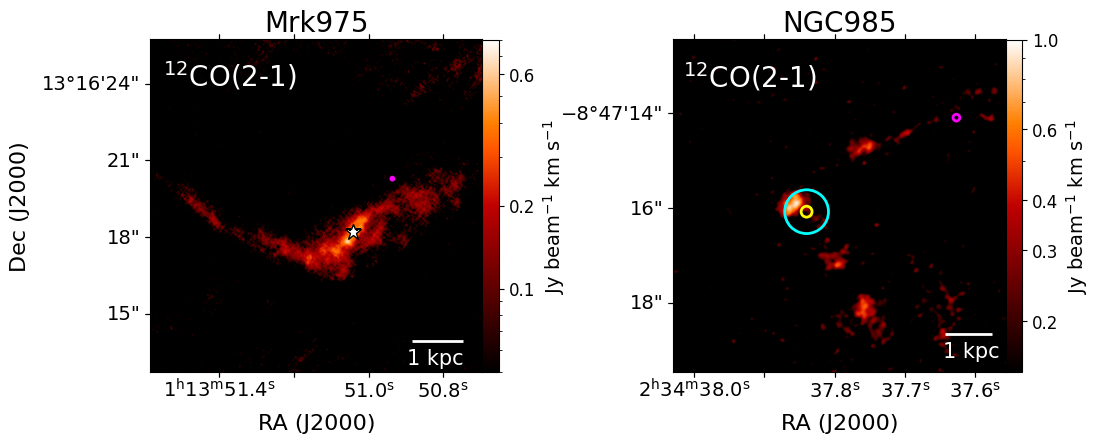}
    \caption{High-resolution $^{12}$CO(2-1) moment 0 (integrated intensity) maps of single AGN in major mergers. For the AGN, the upper-bound and lower-bound estimates of the SMBH SoI are overlaid as cyan and yellow circles, respectively. The white star indicates an unresolved SoI. For the non-AGN SMBH, we mark the upper-bound of the SoI with a magenta circle (lower-bounds were unresolved). Note that the colormap is shown on a logarithmic scale. \label{fig:mom0_single}}
\end{figure*}

Figures \ref{fig:mom0_double} and \ref{fig:mom0_single} show the integrated intensity (moment 0) maps of the ALMA-detected $^{12}$CO emission in the dual AGN and the single AGN, respectively. Note that we analyze $^{12}$CO(1-0) detections for ESO\,253-G003 in this work, whereas we analyze $^{12}$CO(2-1) data for the remaining targets. We produce these maps from the ALMA data cubes by integrating over contiguous channels with $\geq 3\sigma$ detections (velocity ranges of $\pm100$ to $\pm1000$ km s$^{-1}$ of the expected redshifted frequency of the $^{12}$CO emission line). For resolved SoIs of AGN, the upper-bound and lower-bound estimates are overlaid in cyan and yellow, respectively. For systems with unresolved SoIs, we mark the AGN position with a white star. The positions of the AGN were determined from $\sim100$\,GHz or $\sim260$\,GHz continuum detections that will be presented in Droguett et al. (in prep).  These continuum images were extracted from the ALMA data sets that we present in this work, and the astrometry is thus identical to the $^{12}$CO maps. For the non-AGN, the positions were determined from optical centers of the galaxies, reported by the Gaia Space Observatory \citep{Gaia2023}. The upper-bounds of the SoIs are overlaid as magenta circles in  Figure \ref{fig:mom0_single} (lower-bounds are unresolved). 

In some systems, the $^{12}$CO emission traces kpc-scale morphological features of the host galaxy, such as tidal arms (most visible in Mrk\,739 and Mrk\,975) and circumnuclear rings (e.g., Mrk\,463 and Mrk\,739). Other systems like ESO\,253-G003 and NGC\,985 have more spatially compact $^{12}$CO emission features. While the gas morphologies vary substantially in this way, the $^{12}$CO molecular gas is generally concentrated near the AGN position in all galaxies. In the single AGN systems (Mrk\,975 and NGC\,985), however, we find an absence of $^{12}$CO emission near the non-AGN SMBH. We find that these $^{12}$CO non-detections persist even when considering the relative astrometric uncertainty of 0.06" and 0.02" when aligning the Gaia and ALMA coordinates for Mrk 975 and NGC 985, respectively. We determined these uncertainties by comparing the coordinates of the ALMA check calibrator sources to the Gaia catalog. We briefly discuss the absence of $^{12}$CO emission in Section \ref{subsec:singleAGN}.

\begin{figure}
    \centering
    \includegraphics[width=0.49\textwidth]{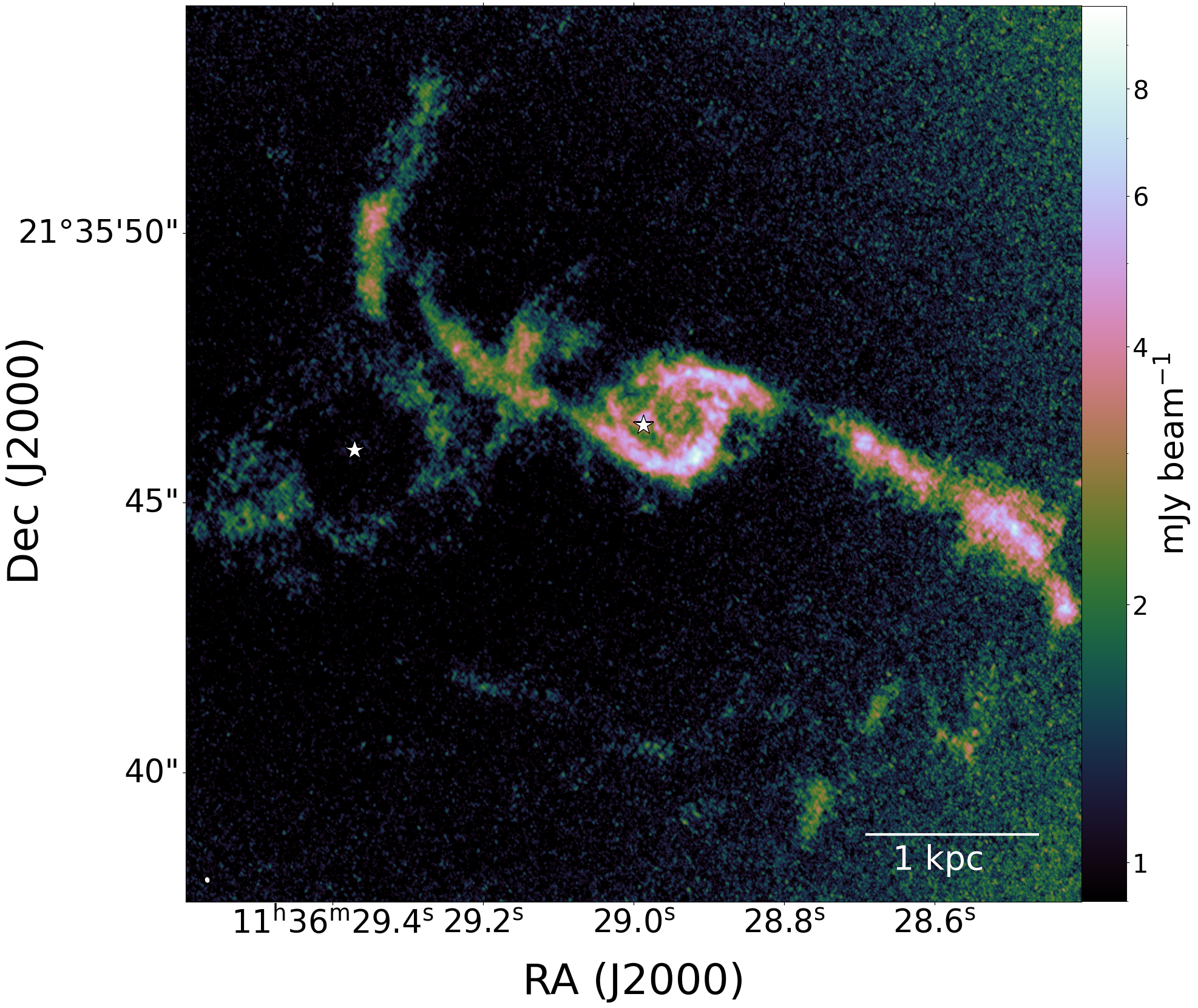}
    \caption{$^{12}$CO(2-1) Moment 8 (spectral line peak) map of Mrk 739. In addition to the prominent $^{12}$CO emission from Mrk\,739W, the circumnuclear ring of Mrk\,739E is faintly visible to the left-hand side. The SMBHs have unresolved SoIs, and we mark their positions with white stars. Note that the colormap is shown on a logarithmic scale. }
    \label{fig:mom8}
\end{figure}

For Mrk\,739, bright $^{12}$CO(2-1) emission is detected near the western nucleus, but the emission is faint surrounding the eastern nucleus. The fainter emission is not visible in Figure \ref{fig:mom0_double} due to the choice of color-scale. To better show the morphology of the emission near this fainter eastern galaxy, we present a moment 8 map (line peak) in Figure \ref{fig:mom8}. We find that the $^{12}$CO(2-1) emission traces a circumnuclear ring with a similar morphology to that of the western nucleus. 

For all galaxies in our sample, we measure the total $^{12}$CO line fluxes from the zeroth moment maps (Figures \ref{fig:mom0_double} and \ref{fig:mom0_single}) by summing the flux densities of all pixels with $>3\sigma$ emission in the 25$''$ ALMA field of view. For Mrk\,463, Mrk\,739, and ESO\,253-G003, we report fluxes for each nucleus separately because the associated $^{12}$CO emission is individually detected and spatially resolved. For the remaining systems, we report a single global measurement. We estimate 1$\sigma$ uncertainties by measuring the RMS in the emission-free regions of the intensity maps and assuming 5\% and 10\% flux uncertainties for ALMA Bands 3 and 6, respectively. These fluxes are reported in Table \ref{tab:CO}. We find that our high-resolution ALMA observations recover $\sim70-80\%$ of the unresolved single-dish flux \citep{evans02, Greve2009, koss11a, koss21}, with the exception of ESO\,253-G003 where we only recover $\sim30\%$ \citep{koss21}, likely due to the lack of TM2 observations. We calculate these percentages by dividing the ALMA-observed global $^{12}$CO flux by the unresolved single dish measurements, after confirming that both observations had spectral coverage greater than the observed line width of the emission line.

Following the prescription described in \citet{solomon05}, we calculate the $^{12}$CO line luminosity of each galaxy,
\begin{equation}
    L^{'}_{\text{CO}} = 3.25\times10^7S_{\text{CO}}\Delta v\, \nu_{\text{obs}}^{-2}\,D_L^2\,(1+z)^{-3}.
    \label{eq:CO_lumin}
\end{equation}
With these line luminosities, we can then obtain the total molecular gas mass of each galaxy by applying the relation,
\begin{equation}
   M_{\text{mol}} = \alpha_{\text{CO}} \frac{L^{'}_{\text{CO(2-1)}}}{r_{21}} = \alpha_{\text{CO}}L^{'}_{\text{CO(1-0)}},
    \label{eq:mol_mass}
\end{equation}
where $r_{21}$ is defined as the ratio between the $^{12}$CO(2-1) and $^{12}$CO(1-0) luminosities, and $\alpha_{\text{CO}}$ is the $^{12}$CO-to-H$_2$ conversion factor. We adopt the median galaxy-integrated $^{12}$CO line ratio $r_{21}\approx 1.09$ from \citet{MA2023}, determined from a sample of 40 U/LIRGs, most of which are gas-rich major galaxy mergers like those in our sample. This is also consistent with previous studies of NGC 6240, a dual AGN system in our sample, which has a mean global $^{12}$CO line ratio of $r_{21} = 1.17 \pm 0.19$ \citep{Cicone2018}. While spatial variations in $r_{21}$ have been observed in some galaxies (e.g., \citealt{Koda2020,Maeda2022}), this value represents our best estimate given that $^{12}$CO line ratios at 50$-$100\,pc resolutions have not been studied in a robust sample of major mergers. Note that since $^{12}$CO(1-0) measurements are available for ESO\,253-G003, we directly measure $M_{\text{mol}}$ from the line luminosity without applying the $r_{21}$ conversion factor. 

For $\alpha_{\text{CO}}$, the typical assumed value for Milky Way-like galaxies is $\alpha_{\text{CO}}$ = 4.6 $M_\odot$ (K km s$^{-1}$ pc$^{-2}$)$^{-1}$ \citep{bolatto13}. More extreme U/LIRGs, however, have significantly lower $\alpha_{\text{CO}}$ values of $\sim0.6-1.8$ $M_\odot$ (K km s$^{-1}$ pc$^{-2}$)$^{-1}$  \citep{downes98, tacconi08, Yamashita2017, HI2019, MA2023}. For the complex merger systems in our sample, it is not appropriate to assume a single global value for $\alpha_{\text{CO}}$ \citep{Scoville2023}, and we thus explore the full range of $^{12}$CO-to-H$_2$ conversion factors, $\alpha_{\text{CO}}$\,=\,$0.6-4.6$ $M_\odot$ (K km s$^{-1}$ pc$^{-2}$)$^{-1}$. The total molecular gas masses calculated at each $\alpha_{\text{CO}}$ value are reported in Table \ref{tab:CO}.

\begin{deluxetable*}{lcccc}[t]
\tablenum{4}
\tablecaption{Physical Properties of the Molecular Gas\label{tab:CO}}
\tablewidth{0pt}
\tablehead{ \colhead{(1)}  & \colhead{(2)} & \colhead{(3)} & \colhead{(4)} & \colhead{(5)} \\
\colhead{System}  & \colhead{$^{12}$CO Flux} & \colhead{$^{12}$CO line Luminosity} & \colhead{Gas Mass ($\alpha_{\text{CO}} = 0.6$)} & \colhead{Gas Mass ($\alpha_{\text{CO}} = 4.6$)} \\[-0.2cm]
 & \colhead{(Jy km s$^{-1}$)} & \colhead{(K km s$^{-1}$ pc$^{−2}$)} & \colhead{log($M_{\text{mol}}/M_\odot$)} & \colhead{log($M_{\text{mol}}/M_\odot$)} 
}
%\decimalcolnumbers
\startdata
\multicolumn{4}{c}{Dual AGN}\\
\hline
NGC 6240 & 1163 $\pm$ 116 & $(8.4 \pm 0.8) \times10^9$ & $9.67 \pm 0.02$ &  $10.6 \pm 0.2$\\
Mrk 463 E  & 9.7 $\pm$ 1.4 & $(3.0 \pm 0.4) \times10^8$ & $8.22 \pm 0.03$ &  $ 9.1 \pm 0.3$\\
Mrk 463 W  & 3.8 $\pm$ 0.9 & $(1.2 \pm 0.3) \times10^8$ & $7.81 \pm 0.05$ &  $ 8.7 \pm 0.4$\\
Mrk 739 E & 2.1 $\pm$ 1.5 & $(2.4 \pm 0.2)\times10^7$ & $7.1\pm0.2$ &  $8.0\pm1.4$\\
Mrk 739 W & 72.4 $\pm$ 7.5 & $(8.3 \pm 0.9)\times10^8$ & $8.66\pm0.02$ &  $9.5\pm0.2$\\
UGC 4211 & 9.1 $\pm$ 1.6 & $(1.2\pm0.2)\times10^8$ & $7.82\pm0.09$ &  $8.7\pm0.3$\\
ESO 253$-$G003 NE & 1.5 $\pm$ 0.5$^*$ & $(1.5\pm0.5)\times10^8$$^*$ & $7.94\pm0.09$ &  $8.8\pm0.7$\\
ESO 253$-$G003 SW & 0.2 $\pm$ 0.3$^*$ & $(2.0\pm3.2)\times10^7$$^*$ & $7.1\pm0.4$ &  $8.0\pm3.0$\\
\hline\hline
\multicolumn{4}{c}{Single AGN}\\
\hline
Mrk 975 & 80.2 $\pm$ 8.4 & $(2.2\pm0.2)\times10^9$ & $9.08\pm0.03$ &  $10.0\pm0.2$ \\
NGC 985 & 1.6 $\pm$ 0.8  & $(3.5\pm1.8)\times10^7$ & $7.3\pm0.1$ &  $8.2\pm0.9$ \\
\enddata
\tablenotetext{*}{For the $^{12}$CO(1-0) transition.}
\tablecomments{Column 1: Name of observed system. Column 2: Total $^{12}$CO fluxes. Uncertainties were calculated using the $1\sigma$ RMS of the data cube and by assuming a 5\% and 10\% flux uncertainty for ALMA Bands 3 and 6, respectively. Column 3: $^{12}$CO line luminosities. Column 4: Total molecular gas mass for CO-to-H$_2$ conversion factor $\alpha_{\text{CO}} = 0.6\, M_\odot$ (K km s$^{-1}$ pc$^{-2}$)$^{-1}$. Column 5: Total molecular gas mass for CO-to-H$_2$ conversion factor $\alpha_{\text{CO}} = 4.6
\,M_\odot$ (K km s$^{-1}$ pc$^{-2}$)$^{-1}$. The measurements in Columns 2 and 3 are for the $^{12}$CO(2-1) line for all targets except ESO\,253-G003, which uses the $^{12}$CO(1-0) line. }
\end{deluxetable*}

\subsection{Kinematics}

We perform kinematic modeling of the ALMA data cubes to determine the rotational component of the $^{12}$CO line emission. This allows us to isolate the non-rotating component of the molecular gas. From the Python tool, KINematic Molecular Simulation (KinMS; \citealt{Davis2013, Davis2020}), we employ the \texttt{KinMS\_fitter}\footnote{\url{https://github.com/TimothyADavis/KinMS\_fitter}} wrapper, which utilizes a Bayesian Markov chain Monte Carlo (MCMC) approach to fit a rotating disk to the data cube. \texttt{KinMS\_fitter} functions by creating a cloud of point-source `cloudlets' and manipulating them to describe the morphology and kinematics of the emission. At least 30,000 iterations were performed for each fitting, with some models having up to 65,000 iterations to obtain reasonable $1\sigma$ errors. 

In the majority of our fittings, we assume a surface brightness with a simple exponential disk profile for each rotating component,
\begin{equation}
    I_{\text{mol}}(r) = I_0e^{\frac{-r}{r_{\text{scale}}}}
\end{equation}
where $I_0$ is the central brightness, and $r_{\text{scale}}$ is the exponential disk scale length. For galaxies with ring-like structures or clumpy emission features, however, we found that the standard analytic models available in \texttt{KinMS\_fitter} (e.g., exponential disks, Gaussians) were not suitable because they assumed a centrally-peaked brightness profile. Instead, we fixed the morphology of these systems from their moment 0 intensity maps (Figures \ref{fig:mom0_double} and \ref{fig:mom0_single}) using the KinMS plugin, \texttt{skySampler}, with a uniform sampling scheme. \texttt{KinMS\_fitter} then scales the surface brightness to model the flux. This was the case for Mrk\,463E, Mrk\,739E, Mrk\,739W, and ESO\,253-G003\,SW. 

For all systems, we assumed an arctangential velocity profile of the form, 
\begin{equation}
    V(r) = \frac{2V_{\text{max}}}{\pi}\arctan\frac{r}{R_{\text{turn}}}
\end{equation}
where $V_{\text{max}}$ is the maximum velocity and $R_{\text{turn}}$ is the turnover radius. While galactic rotation curves are complex, we select this model because arctangential velocity profiles have been shown to provide a good fit to (sub-)kiloparsec $^{12}$CO disks in galaxies across different evolutionary stages (e.g., \citealt{Ruffa2022,Smercina2022,Hogarth2023,Osorno2025}). 

The resulting 3-dimensional model consists of ten free parameters for each rotating disk component: $V_{\text{max}}$, $R_{\text{scale}}$, $R_{\text{turn}}$, the $x$- and $y$-coordinates of the kinematic center, the systemic velocity $V_{\text{sys}}$, the total flux, the velocity dispersion $\sigma$, the position angle $\theta$, and the inclination angle $\varphi$. The maximum and minimum allowed values for each parameter were generally left at KinMS's default settings to avoid introducing biases to the model while still ensuring that the fits were physically meaningful (e.g., maximum velocity cannot be greater than the total velocity range of the data cube, and the inclination must be within 0 and 90 degrees). We did, however, increase the maximum velocity dispersion from 50\,km s$^{-1}$ to 500\,km s$^{-1}$ to account for the highly turbulent nature of these systems. This is motivated by several observational studies of major mergers that found rotating disks with velocity dispersions on orders of 100s of km s$^{-1}$ (e.g., \citealt{DS98, scoville17}). 

Generally, we fit two rotating disks to the $^{12}$CO line emission of each system by defining a spatial region for each component and fitting them separately. For some systems, however, we report single-component models because the $^{12}$CO emission was either undetected near the second nucleus (Mrk\,975) or only one velocity gradient was observed (UGC\,4211). The single velocity gradient in UGC\,4211 was also observed in the Ca II $\lambda$8498, 8542, 8662 stellar absorption lines by \citet{koss23}. In all cases, we examined the model-subtracted residual images to confirm that significant negative features, a common indicator of overfitting or a poor choice of model, are not present.  We also confirm that the fitted components are spatially coincident with the host galaxies' optical disks to ensure that the results are physically meaningful \citep{koss11a, koss18, treister18, Tubin2021}. 

In Figures \ref{fig:kinms_dual} and \ref{fig:kinms_single}, we display the KinMS best-fit models and residual maps for the dual and single AGN systems, respectively. The best-fit parameters for each converged model fit are also listed in Table \ref{tab:mcmc}. Uncertainties are reported at 1$\sigma$ levels. The KinMS-generated corner plots of our models are also available in Appendix \ref{sec:corner}. We find that the MCMC probability distributions for some parameters are bimodal and/or non-Gaussian. This is expected for the complex galaxy mergers studied in this work, given that the molecular gas kinematics are not always dominated by rotation in their turbulent environments. 

Rotating disk models are not reported for NGC 6240 nor NGC 985. For the dual AGN NGC 6240, \cite{treister20} found that the central $^{12}$CO emission is dominated by a central filament-like structure, rather than a rotating disk. To verify this, we attempted to model the emission using both a single-component and a two-component rotating disk model. The best-fit models for both set-ups, however, were inconsistent with expected positions and orientations of the stellar disks \citep{Baan2007, treister20} and are thus unlikely to be physically meaningful. For the two-component model, we also attempted to manually constrain the parameter space (e.g., restricting the possible kinematic centers of each disk component), but the model was unable to converge after 30,000 iterations. We therefore conclude that our results align with that of \cite{treister20}, and the kinematics of NGC 6240 are not dominated by rotation. For the single AGN NGC 985, the velocity profile was similarly inconsistent with that of a rotating disk. Both the single-component and two-component MCMC models failed to converge after 30,000 iterations. For both NGC 6240 and NGC 985, we hereafter assume that all of the molecular gas within the SoIs are non-rotating. 

\begin{deluxetable*}{lcccccccccc}
\tablenum{5}
\tablecaption{Best-fit MCMC Parameters \label{tab:mcmc}}
\tablewidth{0pt}
\tablehead{ \colhead{(1)}  & \colhead{(2)} & \colhead{(3)} & \colhead{(4)} & \colhead{(5)} & \colhead{(6)} & \colhead{(7)} & \colhead{(8)} & \colhead{(9)} & \colhead{(10)} & \colhead{(11)}  \\
\colhead{Component/}  & \colhead{$X_c$} & \colhead{$Y_c$} & \colhead{$V_{\text{sys}}$} & \colhead{Total Flux} & \colhead{$\sigma$} & \colhead{P.A.} & \colhead{Inclination} & \colhead{$R_{\text{scale}}$}  & \colhead{$V_{\text{max}}$} & \colhead{$R_{\text{turn}}$}  \\[-0.2cm]
 \colhead{System} & \colhead{(deg)} & \colhead{(deg)} & \colhead{(km s$^{-1}$)} & \colhead{(Jy km s$^{-1}$)} & \colhead{(km s$^{-1}$)} & \colhead{(deg)} & \colhead{(deg)} & \colhead{(arcsec)} & \colhead{(km s$^{-1}$)} & \colhead{(arcsec)} }
\startdata
\multicolumn{11}{c}{Dual AGN}\\
\hline 
NGC\,6240$^{**}$ & -- & -- & -- & -- & -- & -- & -- & -- &  \\
%NGC 6240 & 253.245435$^{+0.000002}_{-0.000002}$ & 2.401054$^{+0.000002}_{-0.000002}$ & 73$^{+2}_{-2}$ & 282$^{+2}_{-1}$ & 359.9$^{+0.1}_{-0.1}$ & 38$^{+1}_{-1}$ & 1.09$^{+0.01}_{-0.01}$ & 179$^{+7}_{-10}$ & 0.09$^{+0.02}_{-0.04}$\\
Mrk\,436E & 209.012147$^{+0.000001}_{-0.000001}$ & 18.371826$^{+0.000001}_{-0.000001}$ & 70$^{+2}_{-2}$ & 48.9$^{+0.9}_{-0.9}$ & 82$^{+2}_{-2}$ & 24$^{+1}_{-2}$ & 8$^{+4}_{-1}$ & -- & 774$^{+135}_{-241}$ & 0.004$^{+0.009}_{-0.003}$ \\
Mrk\,463W & 209.010857$^{+0.000004}_{-0.000004}$ & 18.371651$^{+0.000003}_{-0.000003}$ & $-$47$^{+4}_{-4}$ & 19.3$^{+0.7}_{-0.7}$ & 100$^{+5}_{-6}$ & 237$^{+2}_{-2}$ & 59$^{+2}_{-2}$ & 0.67$^{+0.02}_{-0.02}$ & 43$^{+7}_{-7}$ & 0.013$^{+0.024}_{-0.009}$\\
Mrk\,739E & 174.122383$^{+0.000007}_{-0.000008}$ & 21.596087$^{+0.000007}_{-0.000006}$ & $-$19$^{+1}_{-1}$ & 32$^{+1}_{-1}$ & 13$^{+0.9}_{-0.7}$ & 115.0$^{+0.5}_{-0.5}$ & 39$^{+1}_{-1}$ & -- & 223$^{+5}_{-4}$ & 0.14$^{+0.04}_{-0.04}$\\
Mrk\,739W & 174.120600$^{+0.000002}_{-0.000002}$ & 21.596270$^{+0.000001}_{-0.000001}$ & 73.1$^{+0.7}_{-0.6}$ & 114$^{+2}_{-2}$ & 33.2$^{+0.5}_{-0.6}$ & 33$^{+1}_{-1}$ & 47$^{+4}_{-5}$ & -- &  54$^{+4}_{-2}$ & 0.02$^{+0.03}_{-0.01}$ \\
UGC\,4211 & 121.193296$^{+0.000004}_{-0.000003}$ & 10.776699$^{+0.000004}_{-0.000005}$ & -44$^{+3}_{-3}$ & 124$^{+3}_{-3}$ & 88$^{+3}_{-2}$ & 18.2$^{+0.8}_{-0.8}$ & 72.3$^{+0.8}_{-0.8}$ & 1.60$^{+0.04}_{-0.08}$ & 254$^{+9}_{-10}$ & 0.26$^{+0.06}_{-0.05}$ \\
ESO\,253-G003 NE$^*$ & 81.325526$^{+0.000002}_{-0.000002}$ & $-$46.005607$^{+0.000002}_{-0.000002}$ &        12$^{+3}_{-3}$  & 6.2$^{+0.4}_{-0.4}$ & 39$^{+3}_{-2}$ & 118$^{+5}_{-5}$ & 30$^{+7}_{-9}$ & 0.15$^{+0.01}_{-0.01}$ & 579$^{+134}_{-166}$ & 0.7$^{+0.3}_{-0.3}$ \\
ESO\,253-G003 SW$^*$ & 81.325028$^{+0.000002}_{-0.000001}$ & $-$46.005869$^{+0.000002}_{-0.000002}$ & 124$^{+13}_{-10}$ & 1.7$^{+0.2}_{-0.2}$ & 86$^{+14}_{-13}$ & 278$^{+7}_{-8}$ & 62$^{+12}_{-17}$ & -- & 360$^{+86}_{-89}$ & 0.09$^{+0.08}_{-0.06}$\\
\hline\hline
\multicolumn{11}{c}{Single AGN}\\
\hline
NGC\,985$^{**}$ & -- & -- & -- & -- & -- & -- & -- & --  \\
Mrk\,975 & 18.462667$^{+0.000003}_{-0.000003}$ & 13.271717$^{+0.000004}_{-0.000003}$ & 43$^{+2}_{-2}$ & 148$^{+3}_{-3}$ & 64$^{+2}_{-1}$ & 55.1$^{+0.5}_{-0.5}$ & 8.1$^{+0.4}_{-0.3}$ & 1.02$^{+0.02}_{-0.02}$ & 1442$^{+40}_{-63}$ & 0.08$^{+0.02}_{-0.01}$\\
\enddata
\tablenotetext{*}{For $^{12}$CO(1-0) transition.}
\tablenotetext{**}{Model is not reported. The velocity profile was inconsistent with that of a rotating disk.}
\tablecomments{Column (1): Name of system and component. Column (2): X-coordinate of kinematic center in degrees. Column (3): Y-coordinate of kinematic center in degrees. Column 4 Systemic velocity of the $^{12}$CO rotating disk in km s$^{-1}$. Here, `0 km s$^{-1}$' corresponds to the systemic velocity of the host galaxy, inferred from the redshifts reported in Table \ref{tab:sample} Column (5): Total flux of the rotational component. Note that KinMS includes pixels with $<3\sigma$ flux densities when determining this quantity, resulting in a significantly greater flux measurement than what is reported in Tables \ref{tab:CO} and \ref{tab:mol_SOI}. Column (6): Velocity dispersion in km s$^{-1}$. Column (7): Position angle of kinematic major axis in degrees. Column (8): Inclination of the rotating disk in degrees. Column (9): Scale radius for exponential disk profile in arcseconds. Note that this parameter was not considered for systems fitted with a fixed surface brightness profile via \texttt{SKySampler}. Column (10): Maximum velocity in km s$^{-1}$. Column (11): Turnover radius of arctangential velocity profile in arcseconds. For all parameters, we report 1$\sigma$ uncertainties.}
\end{deluxetable*}

\begin{figure*}
\begin{tabular}{c}
  \includegraphics[width=\textwidth]{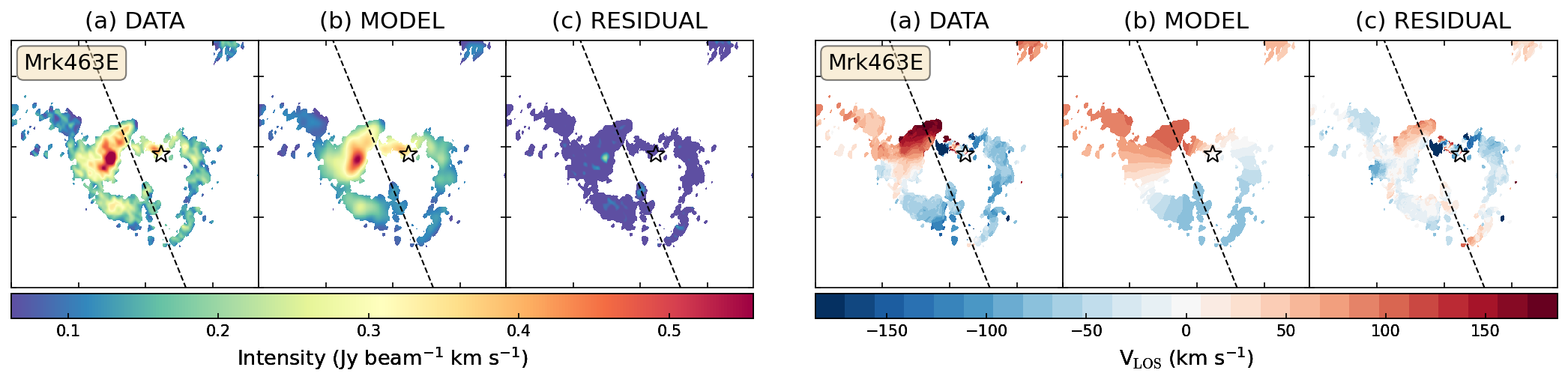} \\
  \includegraphics[width=\textwidth]{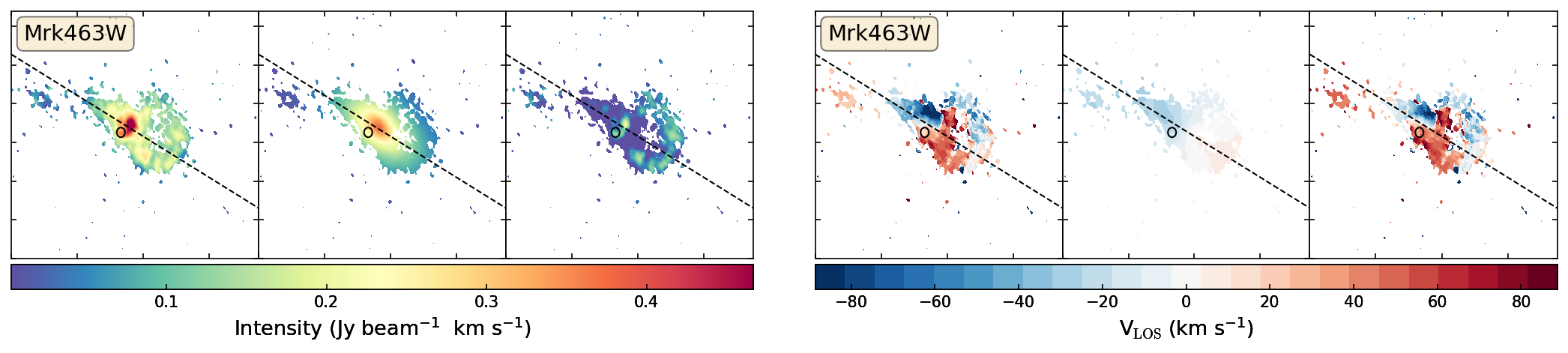} \\
  \includegraphics[width=\textwidth]{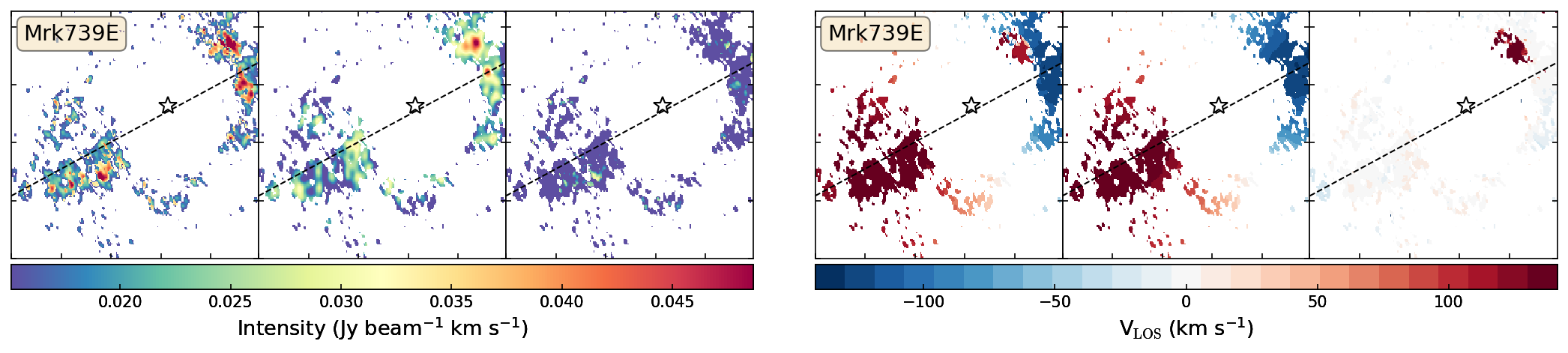} \\
  \includegraphics[width=\textwidth]{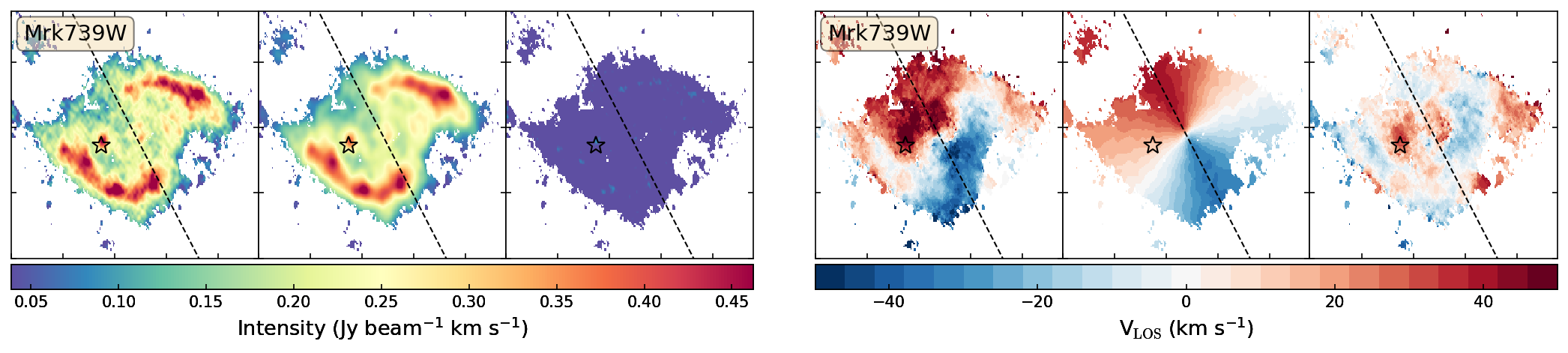} \\
\end{tabular}
\caption{Best-fit models created using KinMS for dual AGN systems. Each row of panels refers to a single system. The left-hand column of panels refers to the zeroth moment (integrated intensity) of the $^{12}$CO data cube, whereas the right-hand column refers to the first moment (velocity profile). From left to right, each column shows the data, model, and residual map. Primarily for visual purposes, the data, model, and residual maps have been masked to spatial regions that had pixels with $\geq3\sigma$ $^{12}$CO detections in the ALMA cube. For Mrk\,739E and ESO\,253-G003\,SW, however, we lower this threshold to $\geq2\sigma$ to account for the faintness of the emission. The best-fit position angle of the rotating disk is shown as a dashed black line. For systems with resolved SoIs, we show the lower and upper bounds of the SoIs as black circles. If only the SoI upper bound is resolved, we only show the upper bound. For systems with unresolved SoIs at both bounds, we mark the AGN position as a star. \label{fig:kinms_dual}}
\end{figure*}

\begin{figure*}
    \ContinuedFloat
    \captionsetup{list=off,format=cont}
\begin{tabular}{c}
  \includegraphics[width=\textwidth]{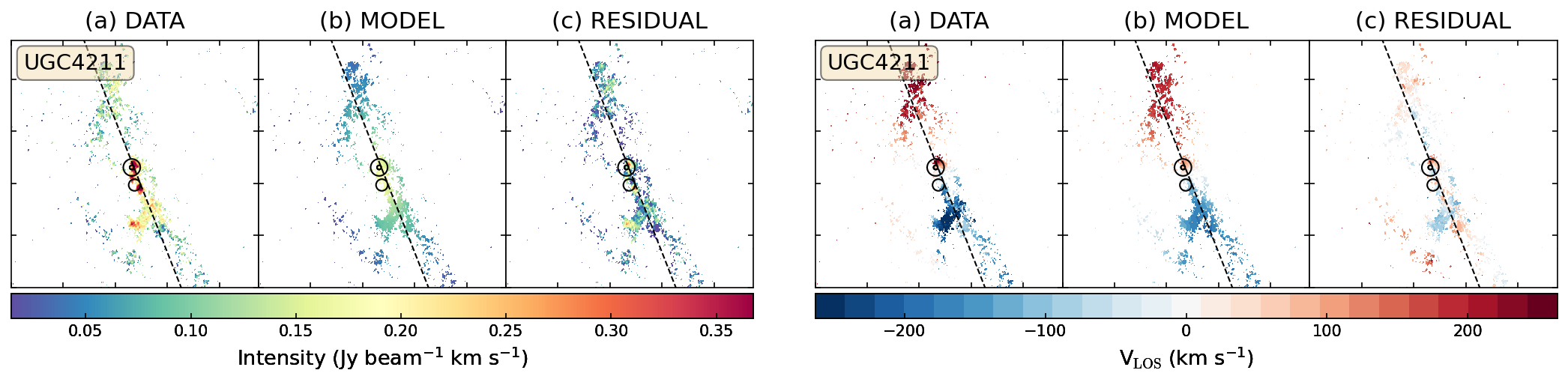} \\
  \includegraphics[width=\textwidth]{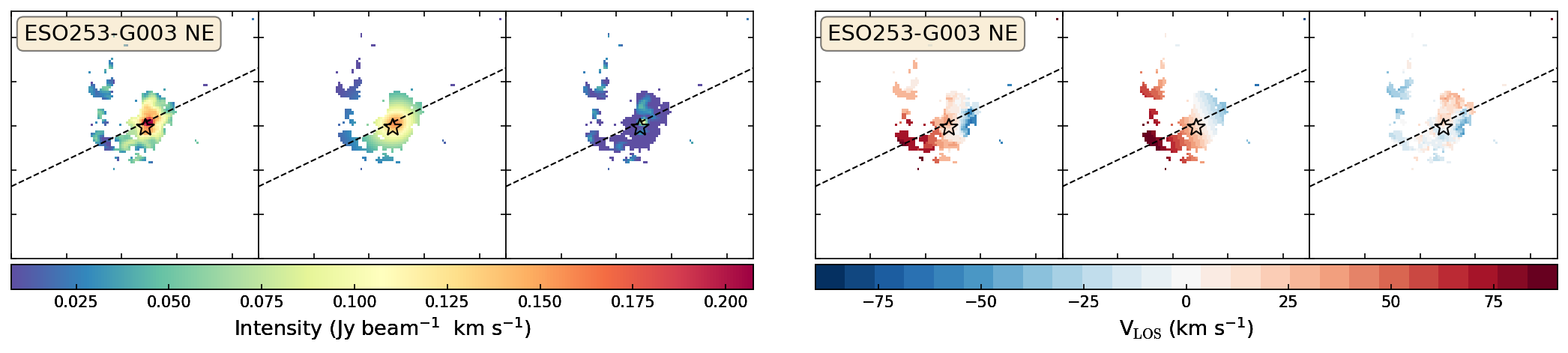} \\
  \includegraphics[width=\textwidth]{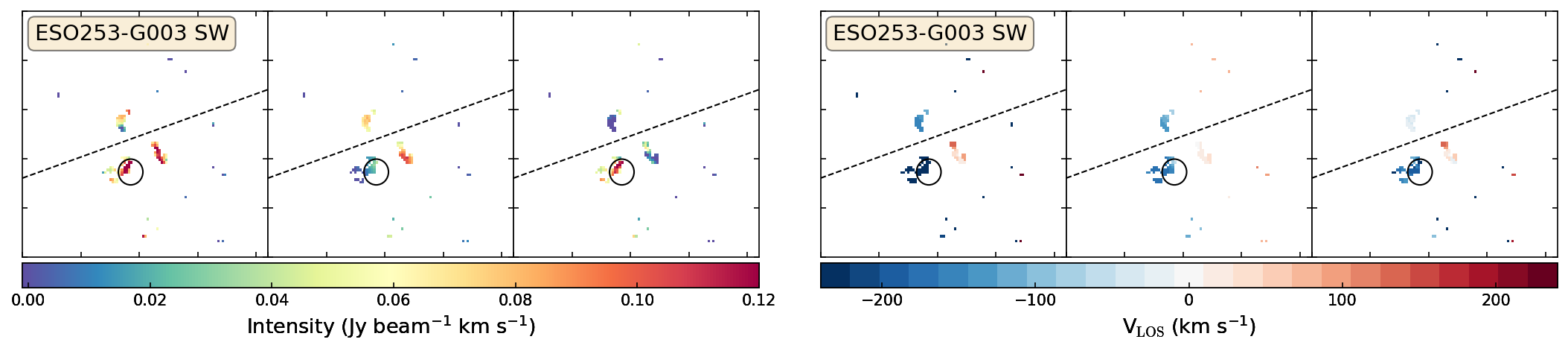}
\end{tabular}
\caption{cont.}
\end{figure*}

\begin{figure*}
\begin{tabular}{c}
  \includegraphics[width=\textwidth]{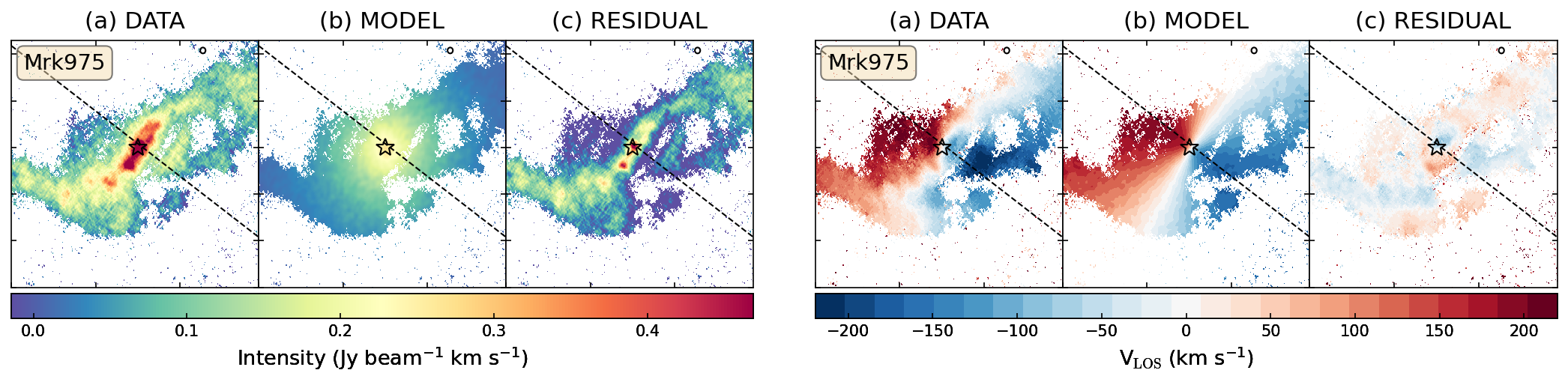} \\
\end{tabular}
\caption{Same as Figure 4 for single AGN. The AGN position is also overlaid as a black star. The upper-bound of non-AGN SMBH's SoI is overlaid as a black circle in the top right. \label{fig:kinms_single}}

% Best-fit models created using KinMS for single AGN systems. Each row of panels refers to a single system. The left-hand column of panels refers to the zeroth moment (integrated intensity) of the $^{12}$CO data cube, whereas the right-hand column refers to the first moment (velocity profile). From left to right, each column shows the data, model, and residual map. Primarily for aesthetic reasons, the data, model, and residual maps have been masked to regions with $\geq3\sigma$ $^{12}$CO detections in the moment 0 map (Figure \ref{fig:mom0_single}). The best-fit position angle of the rotating disk is shown as a dashed black line. The AGN position is also overlaid as a black star. The upper-bound of inactive SMBH's SoI is overlaid as a black circle in the top right. }
\end{figure*}

With these best-fit rotating disk models and the model-subtracted residual images, we can now characterize the molecular gas inside each SoI. Here, the residual images contain only the non-rotating component of the $^{12}$CO molecular gas.  The residual moment 1 maps reveal that the non-rotating gas is predominantly low-velocity (line-of-sight velocity $\lesssim 200$\,km\,s$^{-1}$), and there was no evidence of any significant high velocity components in the any of the residual cubes. To obtain the mass of this low-velocity non-rotating molecular gas from the residual images (third column from the left in Figures \ref{fig:kinms_dual} and \ref{fig:kinms_single}), we extract $^{12}$CO line fluxes from $>3\sigma$ pixels within the SMBH SoIs and derive the molecular gas mass using Equations \ref{eq:CO_lumin} and \ref{eq:mol_mass}, as was done for the global $^{12}$CO emission. We repeat this process for the upper- and lower-bound estimates of the SMBH SoI. In the case of systems with unresolved SMBH SoIs, we extract the $^{12}$CO flux using a beam-sized aperture at the SMBH position and report the molecular gas mass as an upper limit.
Table \ref{tab:mol_SOI} reports the final molecular gas mass measurements\footnote{Formally, these are projected molecular gas masses and should be treated as upper-limits to the true mass within the SMBH SoI.}. 

We emphasize that while our high resolutions resulted in significant flux loss in our global $^{12}$CO emission measurements relative to the single-dish measurements in the literature (as described in Section \ref{subsec:CO_prop}), this missing flux has a negligible impact on the SoI gas masses that we report here. We estimated a maximum impact on the SoI gas measurements by assuming that the lost flux has a smooth distribution across the single-dish primary beam ($\sim20-28”$ diameters). We find that the maximum flux loss within the SoI corresponds to less than the 1$\sigma$ RMS of the ALMA data cubes presented in this work and is thus unlikely to significantly impact the reported molecular gas masses within the SoI. The reported SoI gas fractions (the molecular gas within the SoI divided by the global molecular gas of the host galaxy; columns 3 and 6 in Table \ref{tab:mol_SOI}), however, likely overestimate the true fraction by approximately the same order of magnitude as the flux loss on global/kiloparsec scales (a factor of $\sim3$ for ESO\,253-G003, and a factor of $\sim1.2-1.4$ for the remaining systems).

\begin{deluxetable*}{lccc|ccc}
\tablenum{6}
\tablecaption{Molecular Gas Mass within the Upper and Lower Bound Estimates of the SoI \label{tab:mol_SOI}}
\tablewidth{0pt}
\tablehead{ & \multicolumn{3}{|c|}{SoI Lower-bound} & \multicolumn{3}{c}{SoI Upper-bound} \\
\hline 
\colhead{(1)} & \colhead{(2)} & \colhead{(3)} & \colhead{(4)} & \colhead{(5)} & \colhead{(6)} & \colhead{(7)} \\
\colhead{System} & \colhead{Molecular Gas} & \colhead{Fraction of Global} & \colhead{Non-rotating} & \colhead{Molecular Gas} & \colhead{Fraction of Global} & \colhead{Non-rotating} \\[-0.2cm]
 & \colhead{log($M_{\text{mol}}/M_\odot$)} & \colhead{Molecular Gas} & \colhead{log($M_{\text{mol}}/M_\odot$)}  & \colhead{log($M_{\text{mol}}/M_\odot$)} & \colhead{Molecular Gas} & \colhead{log($M_{\text{mol}}/M_\odot$)} }
%\decimalcolnumbers
\startdata
\multicolumn{7}{c}{Dual AGN}\\
\hline
NGC\,6240 & & & \\
North & 7.30$-$8.18 & 0.9\% & 7.30$-$8.18 & 8.30$-$9.19 & 8.7\% & 8.30$-$9.19 \\
South & 7.93$-$8.81 & 3.7\% & 7.93$-$8.81 & 8.92$-$9.80 & 35.9\% & 8.92$-$9.80\\
Mrk\,463 & & & \\
East & $<$5.89$-$6.76 & $<$0.5\% & $<$4.63$-$5.52 &  $<$5.89$-$6.78 & $<$0.5\% & $<$4.63$-$5.52  \\
West & $<$6.57$-$7.45 & $<$5.8\% & $<$6.39$-$7.28 & 6.57$-$7.46 & 5.8\% & 6.40$-$7.29  \\
Mrk\,739 & & & \\
East & $<$5.28$-$6.16$^{**}$ & $<$1.4\%$^{**}$ & $<$5.28$-$6.16$^{**}$ &   $<$5.28$-$6.16$^{**}$ & $<$1.4\%$^{**}$ & $<$5.28$-$6.16$^{**}$  \\
West & $<$6.23$-$7.12 & $<$0.4\% & $<$5.81$-$6.70 & $<$6.23$-$7.12 & $<$0.4\% & $<$5.81$-$6.70 \\
UGC\,4211 & & & \\
North & 6.21$-$7.10 & 4.9\% & 6.15$-$7.04  & 7.03$-$7.91 & 32.1\% & 7.02$-$7.91 \\
South & $<$4.61$-$5.50 & $<$0.1\% & $<$4.40$-$5.29  & 6.37$-$7.25 & 7.0\% & 6.37$-$7.25 \\
ESO\,253-G003 & & & \\
NE & $<$6.83$-$7.71 & $<$7.7\% & $<$6.23$-$7.12  &  $<$6.83$-$7.71 & $<$7.7\% & $<$6.23$-$7.12  \\
SW & $<$5.50$-$6.38$^{**}$ & $<$2.6\%$^{**}$ & $<$5.50$-$6.38$^{**}$  & 5.73$-$6.62  & 4.5\% & 5.73$-$6.62\\
\hline\hline
\multicolumn{7}{c}{Single AGN}\\
\hline
Mrk\,975 & & & \\
NE & $<$6.79$-$7.68 & $<$0.5\% & $<$6.62$-$7.51  & $<$6.79$-$7.68 & $<$0.52\% & $<$6.62$-$7.21 \\
NW$^{*}$ & $<$5.02$-$5.91$^{**}$ & $<$0.02\%$^{**}$ & $<$5.02$-$5.91$^{**}$  & $<$5.50$-$6.39$^{**}$ & $<$0.05\%$^{**}$ & $<$5.50$-$6.39$^{**}$ \\
NGC\,985 & & & \\
East & 4.84$-$5.72 & 0.4\% & 4.84$-$5.72 & 7.07$-$7.95 & 61.8\% & 7.07$-$7.95\\
West$^{*}$ & $<$5.63$-$6.51$^{**}$ & $<$4.5\%$^{**}$ & $<$5.63$-$6.51$^{**}$  & $<$5.81$-$6.69$^{**}$ & $<$3.4\%$^{**}$ & $<$5.81$-$6.69$^{**}$ \\
\enddata
\tablenotetext{*}{Non-AGN}
\tablenotetext{**}{ 3$\sigma$ upper limits due to non-detections of $^{12}$CO within the SoI. }
\tablecomments{Column (1): Name of system and component. Columns (2) and (5): Projected molecular gas mass within the SMBH SoI. Columns (3) and (6): Fraction of the host galaxy's global molecular gas mass that lies within the SoI (Column (2) or (5) divided by the masses reported in Table \ref{tab:CO}). Columns (4) and (7): Projected non-rotating molecular gas mass within the SoI. Columns (2)-(4) correspond to the SoI lower-bound, whereas Columns (5)-(7) correspond to the SoI upper-bound. Upper-limits indicate unresolved SoIs unless otherwise noted. We report a range of measurements to account for a range of possible $\alpha_{\text{CO}}$ values (0.6$-$4.6 $M_\odot$ (K km s$^{-1}$ pc$^{-2}$)$^{-1}$). }
\end{deluxetable*}

\section{Discussion}
\label{sec:disc}

\subsection{Molecular gas across merger stage}

\begin{figure}
    \centering
    \includegraphics[width=\linewidth]{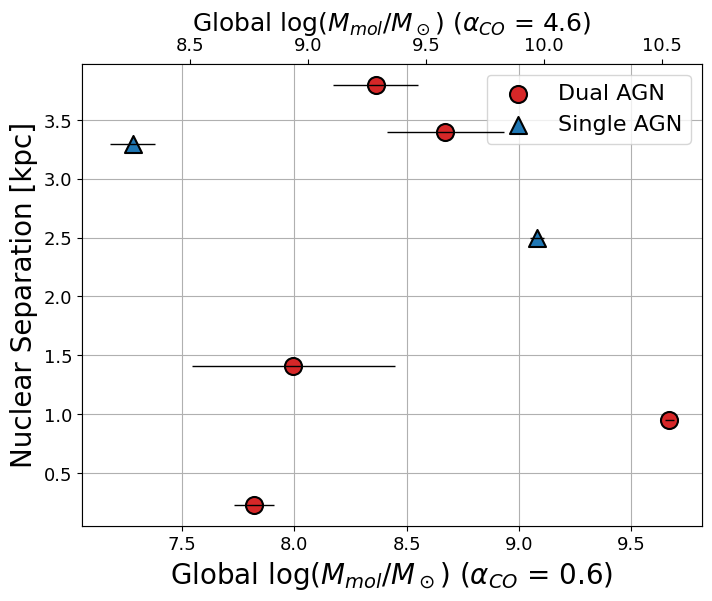}
    \caption{Global molecular gas mass with respect to projected nuclear separation. Gas masses are reported for both $\alpha_{\text{CO}} = 0.6\, M_\odot$ (K km s$^{-1}$ pc$^{-2}$)$^{-1}$ and $\alpha_{\text{CO}} = 4.6\,M_\odot$ (K km s$^{-1}$ pc$^{-2}$)$^{-1}$, but error-bars assume $\alpha_{\text{CO}} = 0.6\,M_\odot$ (K km s$^{-1}$ pc$^{-2}$)$^{-1}$. No significant trend is observed, suggesting that the global supply of molecular gas in the host galaxy does not significantly evolve in the final stages of major mergers.}
    \label{fig:evol}
\end{figure}

We explore the evolution of the host galaxy's total molecular gas content throughout the final stages of the merger sequence. In Figure \ref{fig:evol}, we plot the global molecular gas mass (Table \ref{tab:CO}) of the merger system with respect to their projected nuclear separation (Table \ref{tab:phys_prop}). For systems with spatially separated emission components, we sum the molecular gas masses of the individual components to estimate the global mass. Interestingly, Figure \ref{fig:evol} shows that the gas masses are scattered over $\sim2$ orders of magnitude and does not reveal any significant trend, with a Kendall’s tau correlation coefficient of $\tau =  -0.05$ and a $p$-value of 0.4. This correlation analysis was performed with \texttt{pymccorrelation} \citep{Curran2014, Privon2020}, which calculates the correlation coefficient $\tau$, the $p$-value, and their respective uncertainties using a Monte Carlo approach. The quantities reported here and in the remainder of this work are the median values determined after 1000 bootstrapping iterations. 

We find that at $<$4$\,$kpc projected nuclear separations, the amount of molecular gas in the host galaxy is not dependent on the merger stage. This aligns with that of \citet{Yamashita2017}, who studied the $^{12}$CO(1-0) molecular gas properties of 62 local ($z$$<$0.08) LIRGs and ULIRGs from the Great Observatories All-sky LIRGs Survey (GOALS). This sample is dominated by merging galaxies, similar to those studied in this work. They found that while the spatial scale of the $^{12}$CO(1-0) emission region decreased across the merger sequence, the median molecular gas mass remained constant as the projected nuclear separation decreased \citep{Yamashita2017}. Their findings, combined with the results of our work, display that while merger dynamics may impact the distribution of the global molecular gas, it does not significantly alter the total supply of molecular gas in the central several-kiloparsec region of the galaxy. 

\subsection{Gas inside each SMBH's sphere of influence}
\label{subsec:Gas_SoI}

Our results confirm that major galaxy mergers host large molecular gas reservoirs on orders of $\sim10^5-10^9$ M$_\odot$ (Table \ref{tab:mol_SOI}) within their SMBH SoIs. Here, we consider the potential origins of this molecular gas. 

As nuclear separation decreases, major galaxy interactions reduce the angular momentum of gas, causing inflows that fuel star formation and SMBH growth. These merger-driven mechanisms are thought to be responsible for the build-up of gas and dust in the nuclei of interacting galaxies. Recent studies of AGN obscuration in mergers have provided substantial evidence of this process. For example, column density estimates from \textit{NuSTAR} hard X-ray observations demonstrated that smaller projected nuclear separations correspond to larger columns of obscuring nuclear material \citep{ricci17,Ricci2021}. This is consistent with \cite{kocevski15}, who found that a high fraction ($\sim$65\%) of obscured AGN is hosted by major mergers, implying a direct relationship between merger dynamics and the accumulation of nuclear and circumnuclear material. If the ALMA-observed molecular gas reservoirs within the SoI originated from similar merger-driven mechanisms, we would therefore expect a significant correlation between merger stage and the mass of the molecular gas reservoirs. 

We explore this potential relationship by exploring the correlation between the projected nuclear separation and the amount of molecular gas within the SoI, expressed as both a gas mass and a fraction of the host galaxy's global molecular gas content. In our analysis, we examine the relations using both the lower- and upper-bound estimates of the SoI. We acknowledge, however, that the SoI lower-bound is unresolved in a high fraction of sources, resulting in the molecular gas measurements being dominated by upper-limits. We therefore lack sufficient information to quantify a precise correlation at the SoI lower-bound. For this reason, we emphasize that the upper-bound results, with a higher fraction of resolved measurements, possess a higher statistical significance and will serve as the cornerstone of our discussion hereafter. 

\begin{figure*}
    \begin{subfigure}{0.5\textwidth}
      \includegraphics[width=\textwidth]{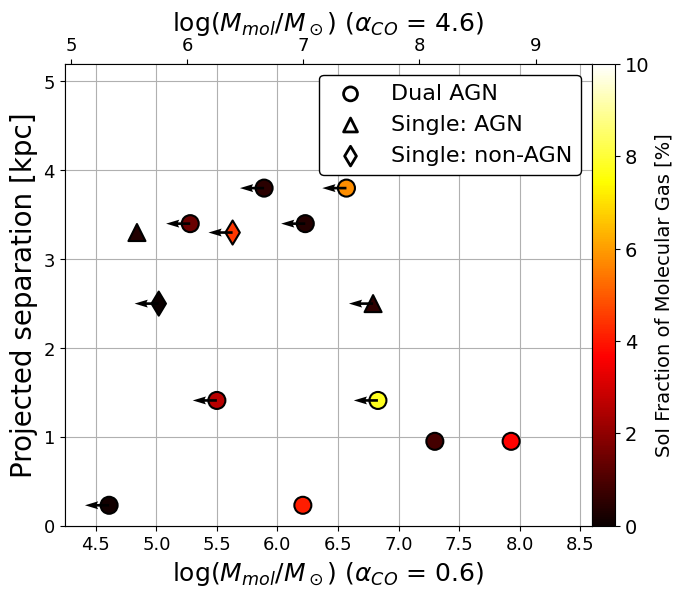}
      \caption{SoI Lower-bound \label{fig:Sep_Frac_Totala}}
    \end{subfigure}
    \hfill
    \begin{subfigure}{0.5\textwidth}
      \includegraphics[width=\textwidth]{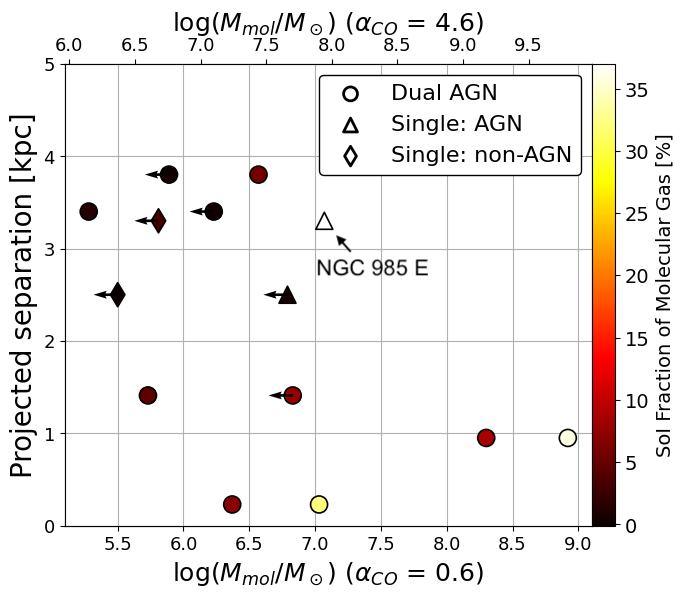}
      \caption{SoI Upper-bound \label{fig:Sep_Frac_Totalb}}
    \end{subfigure}
    \vspace{-0.5cm}
    \caption{Projected nuclear separation with respect to the molecular gas mass ($M_{\text{mol}}$/$M_\odot$) within the SMBH SoI at both the SoI lower-bound (a) and upper-bound (b) estimates. The color scale indicates the fraction of the global molecular gas that lies within the SoI. Upper limits (shown as black arrows) are for unresolved SoI and/or $^{12}$CO non-detections. We find that, at lower projected nuclear separations, a higher fraction of the host galaxy's molecular gas tends to lie within the SMBH SoI. \label{fig:Sep_Frac_Total}}
  \end{figure*}

While we find no obvious trend between the projected nuclear separation and the molecular gas within the SoI at the lower-bound (Figure \ref{fig:Sep_Frac_Totala}), SoI upper-bound measurements show that, except for NGC\,985\,E, smaller projected nuclear separations ($<1.5$ kpc) tend to correspond to molecular gas reservoirs that contain higher fractions of the host galaxy's total molecular gas (Figure \ref{fig:Sep_Frac_Totalb}). In fact, systems with projected nuclear separations $<$1.5\,kpc have a median SoI fraction of 8\%, compared to $<1$\% for projected nuclear separations between 1.5$-$4\,kpc. These trends remain significant even when we consider the uncertainties introduced by the loss of diffuse flux in our global measurements. While this 1.5\,kpc threshold was empirically selected, these differences emphasize that the nuclear molecular gas properties of the host galaxy continue to evolve until the final stages of the merger. Such trends are consistent with merger-driven dynamics funneling the host galaxy's molecular gas toward the nuclear regions and feeding the molecular gas reservoirs probed in this work. Our results indicate that galaxy interactions play a critical role in the nuclear structure of molecular gas in merger systems.  

\subsection{Spatial misalignment with $^{12}$CO peaks}

Examining the distribution of the $^{12}$CO molecular gas, we find that $\sim67\%$ (8/12) of the AGN are spatially offset from their $^{12}$CO emission line peaks (observed in both the $J=2-1$ and $J=1-0$ lines; Figures \ref{fig:mom0_double}, \ref{fig:mom0_single}, and \ref{fig:kinms_dual}). A possible interpretation is that the molecular gas in the highly irradiated environments surrounding the AGN is at a higher excitation than what is traced by the $^{12}$CO(2-1) and $^{12}$CO(1-0) lines studied in this work. As opposed to the far-ultraviolet photons from hot massive stars in photodissociation regions (PDRs), the hard X-ray radiation from an AGN can penetrate higher column densities \citep{Maloney1996, Meijerink2005}, resulting in higher kinetic temperatures of the dense molecular gas.  Higher-$J$ rotational transitions of $^{12}$CO have higher critical densities and excitation temperatures that, therefore, may be better tracers of the molecular gas in the X-ray dominated regions (XDRs) surrounding the AGN \citep{Wolfire2022, Esposito2024}. Future spatially resolved observations of $^{12}$CO line ratios in these regions are necessary to explore this further. 

Alternatively, AGN feedback (e.g., outflows and jets) and SMBH accretion events could be driving the destruction and/or depletion of molecular gas in the vicinity of the AGN. For example, \citet{GB2021}, who studied ten nearby ($D$\,$<$\,28\,Mpc) Seyfert galaxies, found that the ratio of the nuclear-to-circumnuclear molecular gas surface densities (at $r<50$\,pc and $r<200$\,pc, respectively) were inversely correlated with both Eddington ratio and AGN luminosity. In their interpretation, they propose that AGN activity could result in lower circumnuclear molecular gas concentrations. Similar trends have been observed by recent X-ray studies that found that high accretion rates correspond to lower covering factors of circumnuclear material in AGN (e.g., \citealt{ueda03,maiolino07,ricci17b,Ricci2023b}).

The sample analyzed by \citet{GB2021}, however, is dominated by isolated systems with significantly lower concentrations of nuclear gas than the major mergers studied in this work. By the radiation-regulated SMBH growth model \citep{Ricci2022}, radiation pressure efficiency depends on the Eddington ratio and the nuclear column density \citep{Ricci2023b}. The relationship between nuclear gas depletion and AGN activity could thus be different in denser, more extreme environments. Following the methodology of \citet{GB2021}, we briefly test this by comparing the gas concentrations in the nuclei of NGC\,6240, UGC\,4211, and ESO\,253-G003 --- the three systems for which we have the necessary $\sim50$ parsec resolution. 

\begin{figure}
    \centering
    \includegraphics[width=0.5\textwidth]{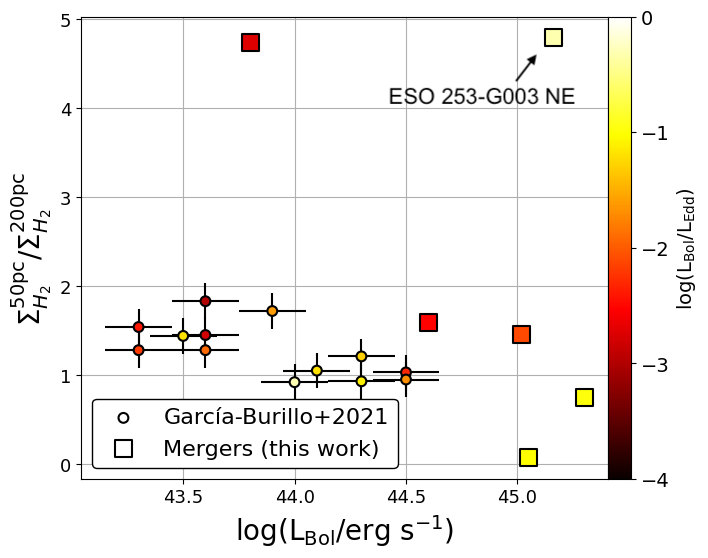}
    \caption{AGN bolometric luminosity (L$_{\text{Bol}}$/erg\,s$^{-1}$) with respect to the ratio of H$_{2}$ surface densities at $r<50$\,pc and $r<200$\,pc ($\Sigma_{H_2}^{\text{50\,pc}}$/$\Sigma_{H_2}^{\text{200\,pc}}$). The colorscale shows the Eddington ratio on a logarithmic scale. This Figure was adapted from \citet{GB2021}. For the surface density ratios, we measure uncertainties assuming $\alpha_{CO}=0.6\,M_\odot$ (K km s$^{-1}$ pc$^{-2}$)$^{-1}$. The mergers generally show an inverse correlation between gas concentration and AGN activity. Due to the small sample size and outlier, however, the potential trend is not statistically significant ($\tau = -0.2$, $p$-value$ = 0.4$). \label{fig:SD_ratio}}
\end{figure}

In Figure \ref{fig:SD_ratio}, we show the ratio of molecular gas surface densities at $r<50$\,pc and $r<200$\,pc with respect to the AGN bolometric luminosity and Eddington ratio. We find that the major mergers studied in this work (plotted as squares) span a wider range of nuclear-to-circumnuclear molecular gas surface density ratios than the low-luminosity Seyfert galaxies (plotted as circles; \citealt{GB2021}). With the exception of ESO\,253-G003 NE, the plotted major mergers follow an inverse correlation between gas concentration and both AGN bolometric luminosity and Eddington ratio, which is qualitatively consistent with the findings of \citet{GB2021}. Due to the small sample size, however, it is unclear if ESO\,253-G003 NE is a mere outlier, and the observed trend is not statistically significant ($\tau = -0.2$, $p$-value = 0.4). High resolution $^{12}$CO observations of a larger sample of AGN in major galaxy mergers are necessary to fully evaluate the observed trend.

\subsection{Off-nuclear SMBHs}

Our disk models (Figures \ref{fig:kinms_dual} and \ref{fig:kinms_single}) also reveal that a high fraction of the observed AGN are spatially offset from the kinematic center of the rotating component of the molecular gas. In Table \ref{tab:offsets}, we report the spatial offsets for each target, omitting NGC 6240 and NGC 985 for which we do not have KinMS models. 

We find that 44\% (4/9) SMBHs have offsets $\gtrapprox$\,200 parsecs, demonstrating the ubiquity of off-nuclear AGN in the observed major mergers. Interestingly, these effects are most apparent in Mrk\,463 and Mrk\,739, which have larger projected nuclear separations ($\sim3-4$\,kpc) compared to the rest of the sample. Here, we do not consider the two nuclei of UGC\,4211 to be offset, given that they are approximately equidistant from the single kinematic center of the molecular gas. 

\begin{deluxetable}{lcc}
\tablecaption{SMBH Offsets from Kinematic Center of Molecular Gas \label{tab:offsets}}
\tablewidth{0pt}
\tablehead{ \colhead{(1)}  & \colhead{(2)} & \colhead{(3)} \\
\colhead{System}  & \colhead{Offset} & \colhead{Offset} \\[-0.2cm]
 & \colhead{($"$)} & \colhead{(pc)} } 
%\decimalcolnumbers
\startdata
\multicolumn{3}{c}{Dual AGN}\\
\hline
Mrk\,463E & 0.174 $\pm$ 0.008 & 194 $\pm$ 9 \\
Mrk\,463W & 0.42 $\pm$ 0.03 & 468 $\pm$ 31 \\
Mrk\,739E & 0.07 $\pm$ 0.04 & 46 $\pm$ 25 \\
Mrk\,739W & 0.616 $\pm$ 0.008 & 416 $\pm$ 5 \\
UGC\,4211N & 0.16 $\pm$ 0.02 & 114 $\pm$ 17 \\
UGC\,4211S & 0.18 $\pm$ 0.02 & 128 $\pm$ 16 \\
ESO\,253-G003 NE & 0.01 $\pm$ 0.01 & 5 $\pm$ 9\\
ESO\,253-G003 SW & 0.22 $\pm$ 0.01 & 196 $\pm$ 10\\
\hline\hline
\multicolumn{3}{c}{Single AGN}\\
\hline
Mrk\,975NE & 0.03 $\pm$ 0.02 & 29 $\pm$ 19\\
\enddata
\tablecomments{Column (1): Name of observed system. Column (2): Spatial offset from kinematic center in arcseconds. Column (3): Spatial offset from kinematic center in parsecs. The uncertainties in columns (2) and (3) were determined from the KinMS $1\sigma$ errors reported in Table \ref{tab:mcmc}. }
\end{deluxetable}

Since SMBH coalescence has not yet occurred in these merger systems, we do not interpret these spatial offsets as evidence of recoiling SMBHs. Instead, gravitational interactions and/or merger-driven tidal torques are likely responsible. Many cosmological simulations of major mergers, however, have fixed the SMBH positions to the gravitational potential minimums of their host galaxies (e.g., \citealt{springel05, Blecha2016}) and are therefore unable to trace the displacement of the SMBHs prior to coalescence. Theoretical models of merger interactions with unconstrained SMBH positions will clarify the nature of the observed spatial offsets.

\subsection{Feeding the Supermassive Black Holes}

The kinematic models of $^{12}$CO emission were able to disentangle the non-rotating and rotating components of the molecular gas within the SMBH SoI (Table \ref{tab:mol_SOI}). Here, we explore the implications of the low-velocity non-rotating gas mass on current and future SMBH growth.\footnote{These non-rotating gas masses serve as upper limits to the total infalling gas mass, given that they may include low velocity outflows, bar-driven streaming, and other components.} 

% Comparisons of molecular gas influx rates and SMBH accretion rates of local major mergers have found that the accretion of molecular gas onto the SMBH is a highly inefficient process \citep{muller-sanchez09, garcia-burillo14, treister18}. Specifically, the infalling gas must lose a significant fraction of its angular momentum prior to accretion, implying that the non-rotating low-velocity (inflowing) component of the molecular gas is the most likely contributor to SMBH growth. The kinematics of the molecular gas within the SMBH SoI could, therefore, provide crucial insight into current and future SMBH growth, which requires observations at very high angular resolutions. 

For example, if we assume all non-rotating molecular gas within the SMBH SoI (see Table \ref{tab:mol_SOI}) to be accreted onto the SMBH eventually, we can roughly estimate the short-term increase in $M_{\text{BH}}$ for each AGN. At the SoI lower-bound, assuming $\alpha_{\text{CO}}=0.6\,M_\odot$ (K km s$^{-1}$ pc$^{-2}$)$^{-1}$, these estimated mass increases range from $0-14$\% and have a median of $\sim$1\%. At the SoI upper-bound, the SMBHs are expected to grow by $\sim0-95\%$ with a median of 3\%. This upper extreme of 95\% corresponds to NGC 6240S and is an outlier likely caused by the fact that the SoI overlaps with a dense molecular gas filament \citep{treister20}. Omitting this outlier, the expected SMBH growth at upper-bounds range from a $\sim0-23\%$ mass increase, with the median remaining at 3\%. These ranges were derived from systems with resolved SoI and do not include the gas mass upper limits presented in Table \ref{tab:mol_SOI}. Given these low medians and wide ranges, we find that while the molecular gas reservoirs are significant, the expected short-term SMBH growth is limited and varies significantly by system.
 
\begin{figure*}
    \begin{subfigure}{0.5\textwidth}
      \includegraphics[width=\textwidth]{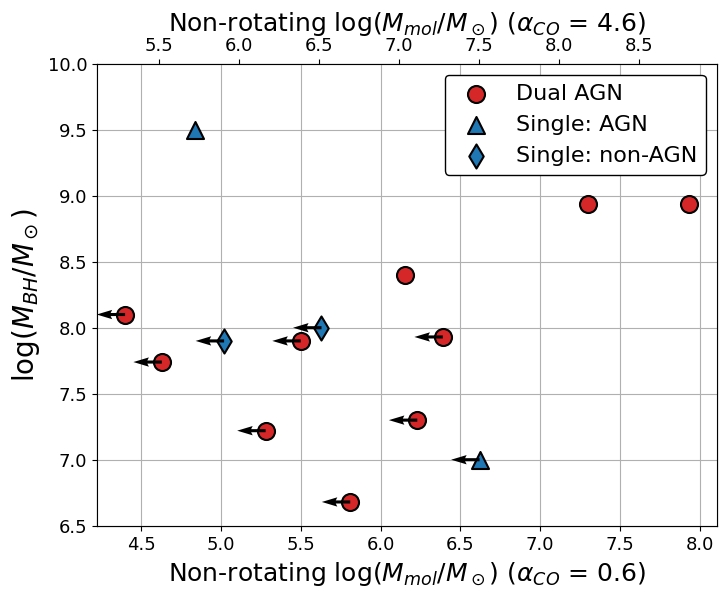}
      \caption{SoI Lower-bound}
    \end{subfigure}
    \hfill
    \begin{subfigure}{0.5\textwidth}
      \includegraphics[width=\textwidth]{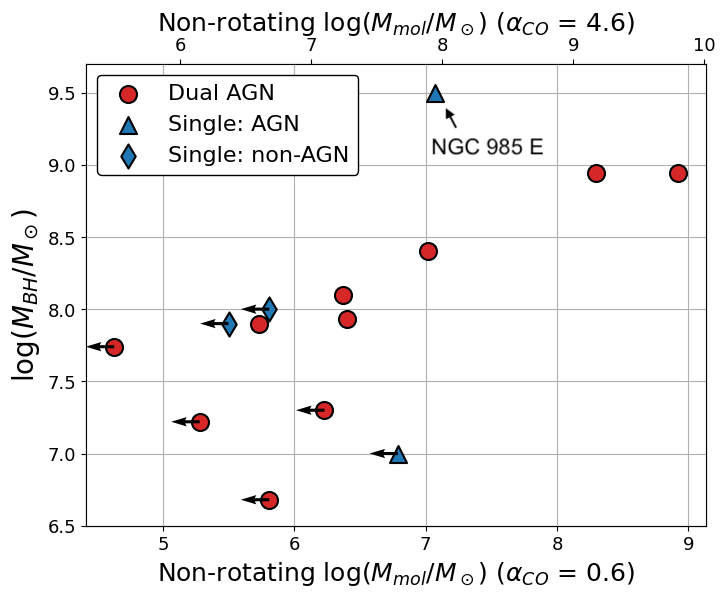}
      \caption{SoI Upper-bound}
    \end{subfigure}
    \vspace{-0.5cm}
    \caption{$M_{\text{BH}}$ with respect to non-rotating molecular gas mass ($M_{\text{mol}}$/$M_\odot$) within the SMBH SoI at both the SoI lower-bound (a) and upper-bound (b). Both axes are on a logarithmic scale. Upper limits (shown as black arrows) are for unresolved SoI and/or $^{12}$CO non-detections.  We find that larger SMBH masses correlate with larger molecular gas reservoirs within the SoI. \label{fig:SMBH_NR}}
\end{figure*}

In addition to these approximate estimates of future SMBH growth, we further explore the link between SMBH accretion and non-rotating molecular gas by comparing the AGN properties of our targets obtained from literature (Table \ref{tab:phys_prop}) to our measurements within the SoI (Table \ref{tab:mol_SOI}). We find that more massive reservoirs of non-rotating molecular gas correlate with higher SMBH masses (Figure \ref{fig:SMBH_NR}). The single AGN NGC\,985E, however, lies above this correlation by approximately an order of magnitude. The sample has a Kendall's tau correlation coefficient of $\tau = 0.65$ and a $p$-value of 0.001. Note that Kendall's tau method incorporates upper limit measurements that are hence included in these calculations. 

\begin{figure*}
    \begin{subfigure}{0.5\textwidth}
      \includegraphics[width=\textwidth]{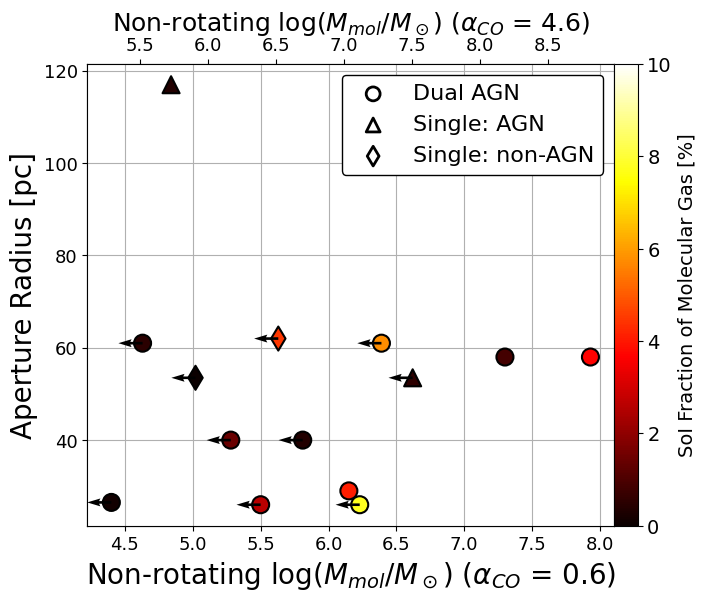}
      \caption{SoI Lower-bound}
    \end{subfigure}
    \hfill
    \begin{subfigure}{0.5\textwidth}
      \includegraphics[width=\textwidth]{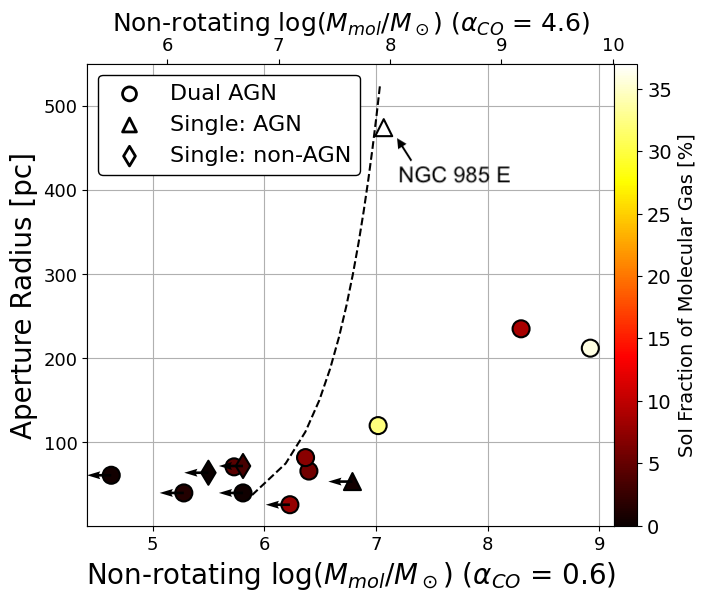}
      \caption{SoI Upper-bound}
    \end{subfigure}
    \vspace{-0.5cm}
    \caption{Aperture radius with respect to non-rotating molecular gas mass ($M_{\text{mol}}$/$M_\odot$) at both the SoI lower-bound (a) and upper-bound (b). Both axes are on a logarithmic scale. Upper limits (shown as black arrows) are for unresolved SoI and/or $^{12}$CO non-detections. The black dashed line indicates a uniform 3$\sigma$ mass distribution. No significant trend is observed, demonstrating that the $M_{\text{BH}}$ correlation in Figure \ref{fig:SMBH_NR} is not a product of differing aperture sizes.  \label{fig:Aper_NR}}
\end{figure*}

\begin{figure*}
    \begin{subfigure}{0.5\textwidth}
      \includegraphics[width=\textwidth]{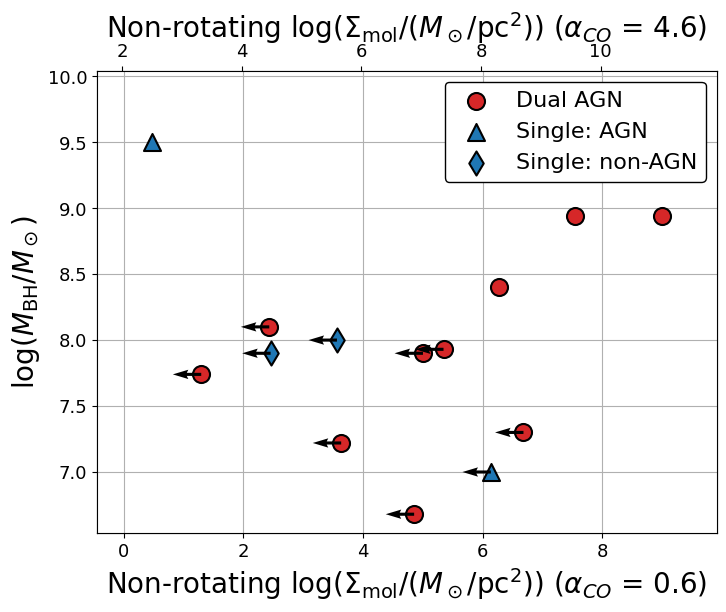}
      \caption{SoI Lower-bound}
    \end{subfigure}
    \hfill
    \begin{subfigure}{0.5\textwidth}
      \includegraphics[width=\textwidth]{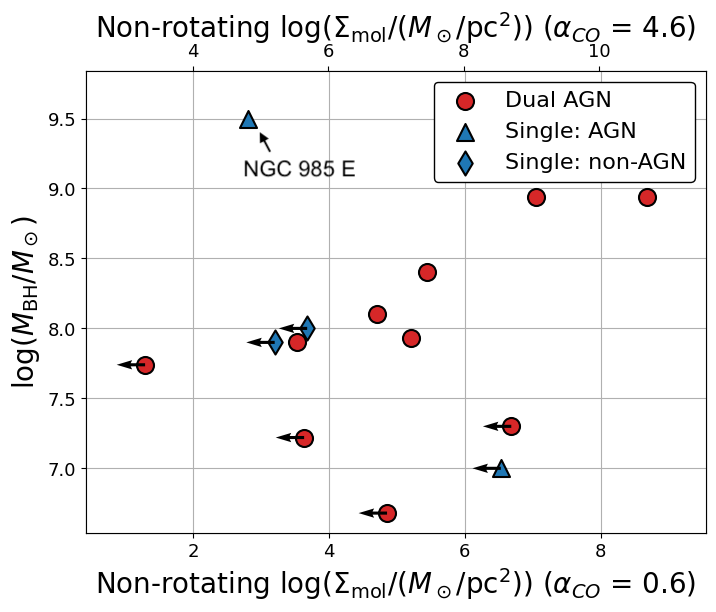}
      \caption{SoI Upper-bound}
    \end{subfigure}
    \vspace{-0.5cm}
    \caption{Surface Density of non-rotating molecular gas mass ($\Sigma_{\text{mol}}/M_\odot/$pc$^2$)) with respect to $M_{\text{BH}}$ at both the SoI lower-bound (a) and upper-bound (b). Both axes are on a logarithmic scale. For systems where the SMBH SoI was smaller than the ALMA data cube's beam size, we measure the molecular gas mass using a beam-size aperture and present the value as an upper limit (shown as black arrows). We find that more massive SMBHs have higher surface densities of non-rotating molecular gas in their SoIs. \label{fig:MBH_SD}}
\end{figure*}

While we considered that the positive trend between $M_{\text{BH}}$ and non-rotating molecular gas mass could be an artifact of differing SoI's (i.e., different aperture sizes used to extract the $^{12}$CO flux), this is unlikely to be the case for our sample. In Figure \ref{fig:Aper_NR}, we plot the aperture radius with respect to the non-rotating molecular gas mass. The plotted dashed line represents a uniform $3\sigma$ mass distribution, taking the median properties of the observed sample (e.g., beam size, RMS noise, luminosity distance). At the SoI upper-bound, we find that the SoI of several nuclei have higher non-rotating molecular gas masses than what could be explained by a uniform mass distribution over a larger spatial scale. This suggests that the $M_{\text{BH}}$ correlation in Figure \ref{fig:SMBH_NR} is driven by differences in molecular gas concentration rather than an increase in spatial scale/aperture size. We confirm this in Figure \ref{fig:MBH_SD} where we observe a correlation between the $M_{\text{BH}}$ and the surface densities of the non-rotating molecular gas within the SoI ($\tau = 0.5$ and $p$-value = 0.02). Omitting the outlier NGC \,985\,E, the correlation significance becomes $\tau = 0.6$ and $p$-value = 0.003. 

Finally, we further confirmed that the observed correlation was not confounded by $M_{\text{BH}}$-dependent properties of the host galaxy by performing partial correlations using the Python package \texttt{Pingouin} to determine Spearman correlation coefficients for the resolved measurements in our sample. We find that even when the stellar velocity dispersion $\sigma_*$ and stellar mass $M_*$ are set as covariates, we observe a strong correlation between the non-rotating molecular gas surface densities and the $M_{\text{BH}}$ (Spearman correlation coefficient $r = 0.8$). We therefore conclude that the nuclear distribution of non-rotating molecular gas is linked to $M_{\text{BH}}$.

% We therefore find that the nuclear distribution of non-rotating molecular gas is linked to $M_{\text{BH}}$ (or a $M_{\text{BH}}$-dependent property of the host-galaxy).

% At the SoI upper-bound, we see potential effects of aperture size when the SoI is large ($r>100$ pc), but the curve is relatively flat when the aperture radius is $<100$ pc. Since most of our sample has an aperture radius $<100$ pc, our results are unlikely to be biased by the spatial scale of the flux measurements. Instead, the $M_{\text{BH}}$ correlation in Figure \ref{fig:SMBH_NR} appears to be driven by differences in molecular gas concentration rather than an increase in spatial scale/aperture size. This is confirmed in Figure \ref{fig:MBH_SD} where we observe a moderate correlation ($p$-value of 0.04) between the $M_{\text{BH}}$ and the surface densities of the non-rotating molecular gas.

\begin{figure*}
    \begin{subfigure}{0.5\textwidth}
      \includegraphics[width=\textwidth]{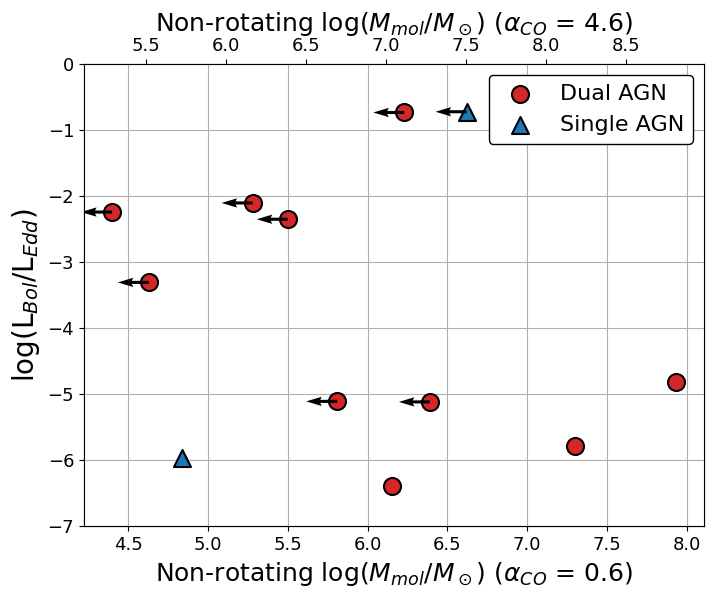}
      \caption{SoI Lower-bound}
    \end{subfigure}
    \hfill
    \begin{subfigure}{0.5\textwidth}
      \includegraphics[width=\textwidth]{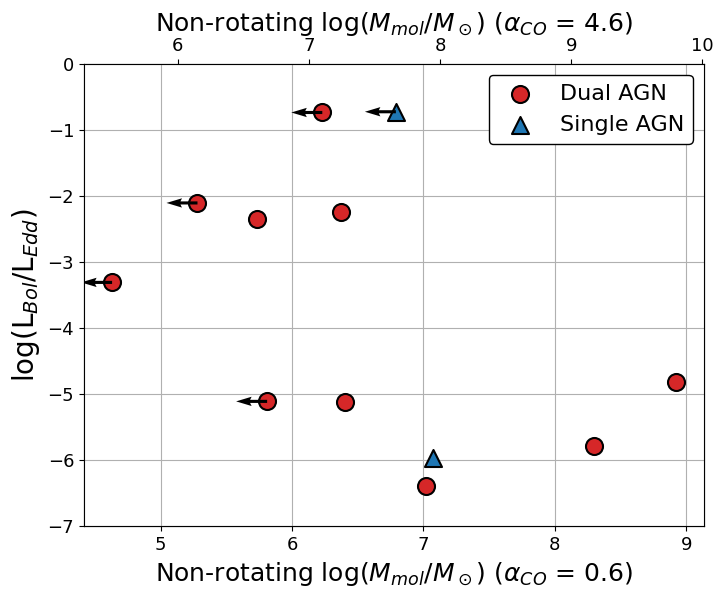}
      \caption{SoI Upper-bound}
    \end{subfigure}
    \vspace{-0.5cm}
    \caption{Eddington ratio (L$_{\text{Bol}}$/L$_{\text{Edd}}$) with respect to the non-rotating molecular gas mass ($M_{\text{mol}}$/$M_\odot$) within the SMBH SoI at both the SoI lower-bound (a) and upper-bound (b). Both axes are on a logarithmic scale. Upper limits (shown as black arrows) are for unresolved SoI and/or $^{12}$CO non-detections.  %We find that lower Eddington ratios moderately correlate with larger non-rotating molecular gas masses within the SMBH SoI.     
    \label{fig:Edd_NR}}
\end{figure*}

\begin{figure*}
    \centering
    \begin{tabular}{cc}
     \includegraphics[width=0.5\linewidth]{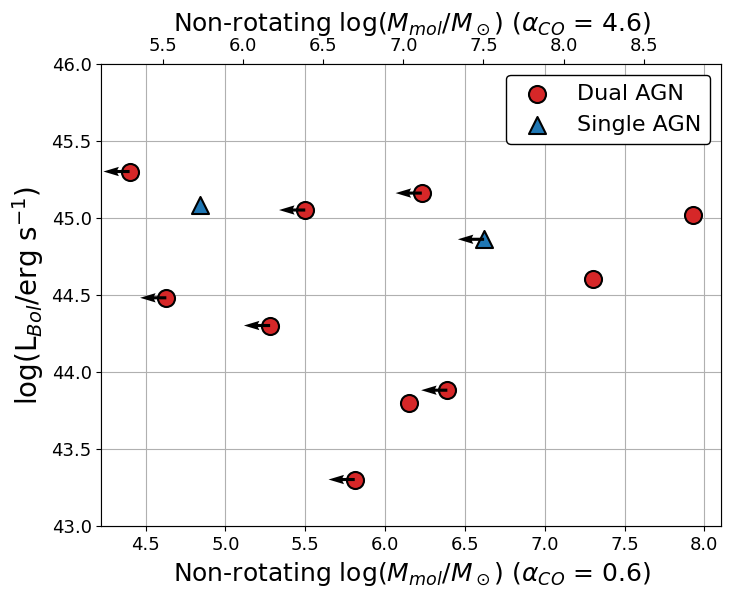}    &  \includegraphics[width=0.5\linewidth]{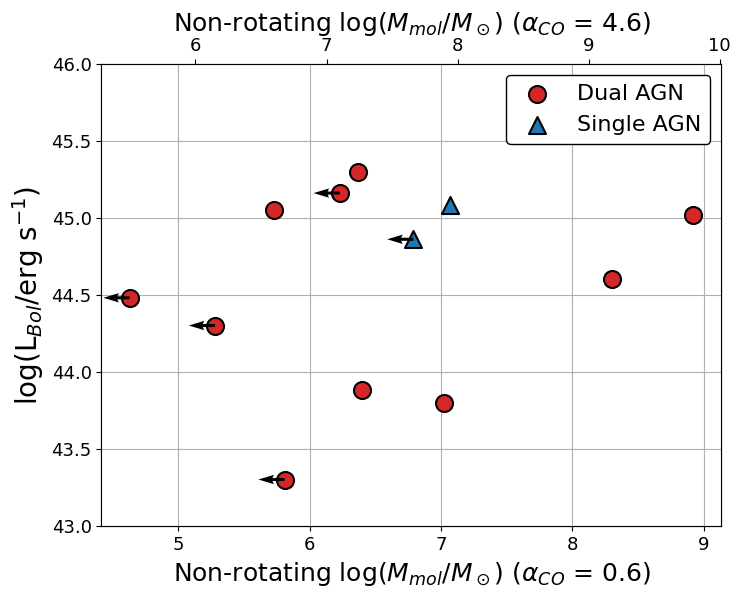}\\
    \end{tabular}
    \caption{Bolometric Luminosity (L$_{\text{Bol}}$) with respect to the non-rotating molecular gas mass ($M_{\text{mol}}$/$M_\odot$)within the SMBH SoI at both the SoI lower-bound (left panel) and upper-bound (right panel). Both axes are on a logarithmic scale. Upper limits (shown as black arrows) are for unresolved SoI and/or $^{12}$CO non-detections. We find no significant correlation.}
    \label{fig:Lumin_NR}
\end{figure*}

Though more massive SMBHs host denser reservoirs of non-rotating molecular gas, we find no obvious relationship between the non-rotating molecular gas mass and the Eddington ratio (Figure \ref{fig:Edd_NR}; $\tau = -0.4$ and $p$-value = 0.05) nor the AGN luminosity (Figure \ref{fig:Lumin_NR}; $\tau = -0.08$ and $p$-value = 0.5). This shows that the availability of non-rotating gas does not elevate the current accretion efficiency of the SMBH, suggesting that only a fraction of the observed non-rotating gas is presently reaching the SMBH. Recent studies of AGN at low redshifts have similarly reported a lack of correlation between AGN luminosity and nuclear molecular gas mass at both high ($\sim$100 pc; \citealt{Elford2024}) and low spatial resolutions (5$-$21 kpc; \citealt{Molina2023}). Our results, combined with the non-correlations from the literature, suggest that the level of nuclear activity in a given merging galaxy cannot purely depend on the amount of non-rotating molecular gas within the SMBH SoI. 

\subsection{Structural differences between Single and Dual AGN} \label{subsec:singleAGN}

A critical question is whether there are intrinsic differences in the molecular gas properties of single and dual AGN systems. A complete sample of single AGN systems will be analyzed in Ramos et al. (in prep.) to perform a comprehensive comparative analysis, but we explore the tentative differences observed in our ALMA data here. 

We find that the non-AGN SMBHs in the single AGN systems contain $\sim$0\% of the global molecular gas of the galaxy (Table \ref{tab:mol_SOI}). Given that accretion onto the SMBH likely requires an influx of molecular gas, it is expected that SMBH inactivity correlates with smaller molecular gas reservoirs in this way. Intriguingly, some dual AGN systems in our sample have SMBHs with similarly small ($<1\%$) molecular gas fractions within the SoI (e.g., Mrk 463E, Mrk 739E; Table \ref{tab:mol_SOI}), suggesting that AGN activity can be present in regions with molecular gas deficits. We therefore find that a limited supply of molecular gas cannot be the only factor that regulates AGN activity.

In comparing the properties of the AGN in each system, we find that the single AGN NGC\,985E is a consistent outlier (Figures \ref{fig:Sep_Frac_Total} to \ref{fig:MBH_SD}). This is particularly notable in Figure \ref{fig:Sep_Frac_Totalb}: the upper-bound SoI of the AGN NGC 985 East contains an exceptionally high fraction of the host galaxy's global molecular gas ($\sim$62\%). This is likely due to fact that, compared to the other systems in our sample, NGC\,985E has a significantly higher estimated SMBH mass and a larger SoI radius ($\sim500$ pc vs. $\sim$15$-$235 pc) despite having a comparable mass surface density within the SoI ($\sim10^3$ M$_\odot$/pc$^{2}$ assuming $\alpha_{CO} = 0.6\, M_\odot$ (K km s$^{-1}$ pc$^{-2}$)$^{-1}$). The larger SoI results in a larger spatial region probed in our measurements of NGC\,985, making it reasonable that the derived molecular gas fraction was significantly higher than the other systems. 

This high fraction, however, is in drastic contrast with the low SoI fraction of the non-AGN SMBHs, NGC\,985W ($<$0.05\%). NGC\,985 is also unique in that it has a remarkably high mass ratio ($\sim$6.9, Table \ref{tab:phys_prop}) and its estimated SMBH masses differ by over a magnitude (log($M_{\text{BH}}/M_\odot$) = 8.0 vs. 9.5). We speculate that these extreme mass differences could also be the driver of the imbalanced distribution of molecular gas in NGC\,985. The gravitational potential well of the merger system will likely favor the more massive SMBH, causing a higher fraction of the galaxy's molecular gas to be funneled towards that nucleus. Such processes could then cause a deficit of non-rotating/infalling molecular gas surrounding the less-massive SMBH, resulting in a non-active SMBH and preventing the dual AGN phenomenon from occurring. This theory is supported by both recent observations \citep{Shields2012,Stemo2021,Glikman2023} and simulations \citep{blecha13, Steinborn2016, Capelo2017} that predict that dual AGN are more likely to form when the nuclei have similar SMBH masses.

We acknowledge, however, that the properties of Mrk\,975 (the other single AGN system studied in this work) are inconsistent with this theory. Despite having SMBH masses that differ by approximately a magnitude (log($M_{\text{BH}}/M_\odot$) = 7.0 vs. 7.9), AGN activity is only detected for the \textit{less} massive SMBH (Mrk\,975\,NE). The host galaxy's mass ratio is also comparable to those of the dual AGN systems, and thus an imbalanced gravitational potential well cannot solely explain why only one SMBH is active. Given that no other significant deviations in SMBH/AGN properties were observed, we speculate that the differences between single and dual AGN systems are more likely to be driven by AGN variability rather than by an intrinsic difference in the available supply of molecular gas. Future analysis of the molecular gas properties of a larger sample of single AGN systems in major mergers is necessary to confirm this theory.

% We acknowledge, however, that the properties of NGC 985 (East and West) do not deviate from those of dual AGN systems in any other trend observed in this Section. Furthermore, the other single AGN system studied in this work, Mrk 975, has similar SMBH masses and a mass ratio comparable to those of the dual AGN systems. Thus, an imbalanced gravitational potential well cannot solely explain why only one SMBH is active. Given that no other significant deviations were observed, we find that the differences between single and dual AGN are more likely to be driven by AGN variability rather than by an intrinsic difference in the available supply of molecular gas. 

% Fraction of total gas inside sphere of influence as a function of nuclear separation separately for dual and single AGN

\section{Conclusions}
\label{sec:conclusion}

We study the molecular gas properties of single and dual AGN systems in major mergers using high resolution ($\lesssim100$\,pc) ALMA observations of $^{12}$CO(2-1) and $^{12}$CO(1-0). In addition to the host galaxy's global molecular gas, we modeled each system's gas kinematics, measuring the mass of the total and non-rotating components of the molecular gas within the SMBH SoI. Our findings are summarized below:

\begin{enumerate}
    \item Although the global molecular gas mass of the host galaxy is relatively constant across projected nuclear separation in the final stages of the merger ($<4$\,kpc), the SMBH SoI contains higher fractions of molecular gas at smaller nuclear separations. In agreement with previous works, we thus find that the gas distribution becomes more centrally concentrated as the merger sequence progresses. In fact, the median fraction of the host galaxy's molecular gas increases from $<1\%$ to $\sim8\%$ when the projected nuclear separations are less than $<1.5$\,kpc.
    
    %we conclude that while merger dynamics do not directly influence the total molecular gas content of the host galaxy, 
    
    \item We find that $\sim67$\% (8/12) of the AGN are spatially offset from $^{12}$CO emission line peaks. The molecular gas in the vicinity of the AGN is either being destroyed by AGN feedback, depleted by nuclear outflows, accretion onto the SMBH, or it is at a higher excitation than what is traced by the $^{12}$CO(2-1) and $^{12}$CO(1-0) lines. 

    \item We find that $\sim44\%$ (4/9) of the AGN are spatially offset by $\gtrapprox200$\,pc from the kinematic centers of the molecular gas. Merger-driven gravitational interactions could cause the SMBHs to become displaced from their galaxy's centers, even prior to SMBH coalescence.
    
    \item The nuclear distribution of the non-rotating molecular gas appears to be regulated by the SMBH mass. More massive SMBHs tend to have higher surface densities of non-rotating molecular gas within their SoIs. 

    \item The non-rotating molecular gas mass is weakly inversely correlated with the Eddington ratio but does not correlate with AGN bolometric luminosity. The lack of significant trends with tracers of AGN activity suggests that the supply of non-rotating molecular gas in the SMBH SoI does not significantly affect the current accretion efficiency of SMBHs in major mergers. 
    
    \item The molecular gas properties of single and dual AGN systems are relatively consistent. We therefore propose that these systems are not intrinsically different but are, instead, products of AGN variability.
\end{enumerate}

While this work presents a first-of-its-kind high-resolution analysis of the molecular gas properties of dual AGN systems, the number of confirmed close-separation ($<10$\,kpc) dual AGN systems is limited. Further multi-wavelength searches for dual AGN in the local Universe are necessary to expand the sample and obtain more robust statistics on their molecular gas properties. In addition to a comprehensive control sample of single AGN systems in mergers, comparisons to non-AGN mergers may also provide insight into the role of molecular gas in triggering AGN activity. 

\begin{acknowledgments}
We thank the anonymous referee for their very useful suggestions that, in our opinion, significantly improved the quality of this article. This paper makes use of the following ALMA data: ADS/JAO.ALMA\#2015.1.00370.S, \#2021.1.01019.S, \#2022.1.01348.S, and \#2023.1.01196.S. ALMA is a partnership of ESO (representing its member states), NSF (USA) and NINS (Japan), together with NRC (Canada), MOST and ASIAA (Taiwan), and KASI (Republic of Korea), in cooperation with the Republic of Chile. The Joint ALMA Observatory is operated by ESO, AUI/NRAO and NAOJ. The National Radio Astronomy Observatory is a facility of the National Science Foundation operated under cooperative agreement by Associated Universities, Inc.  M.A.J. gratefully acknowledges financial support for this research by the Fulbright U.S. Student Program, which is sponsored by the U.S. Department of State and US-Chile Fulbright Commission. Its contents are solely the responsibility of the author and do not necessarily represent the official views of the Fulbright Program, the Government of the United States, or the US-Chile Fulbright Commission.  ET would like to thank the hospitality of the North American ALMA Science Center (NAASC) at NRAO during his sabbatical stay in 2022, where a significant fraction of this work was carried out. AMM acknowledges support from the NASA Astrophysics Data Analysis Program (ADAP) grant number 80NSSC23K0750, from NSF AAG grant \#2009416 and NSF CAREER grant \#2239807, and from the Research Corporation for Science Advancement (RCSA) through the Cottrell Scholars Award CS-CSA-2024-092. We also acknowledge support from: NASA through ADAP awards NNH16CT03C (MK), 80NSSC19K1096 (F.M-S) and 80NSSC23K1529 (F.M-S), ANID through Millennium Science Initiative Program - NCN19\_058 (ET), and ICN12\_009 (FEB), CATA-BASAL - ACE210002 (ET, FEB) and FB210003 (ET, FEB, CR), FONDECYT Regular - 1190818, 1200495, 1241005, and  1250821 (ET, FEB), FONDECYT Postdoctorado 3230653 (IMC) Fondecyt Iniciacion 11190831 (CR), the China-Chile joint research fund (CR), and the European Union’s HE ERC Starting Grant No. 101040227 - WINGS (GV).
\end{acknowledgments}

\vspace{5mm}
\facilities{HST(STIS), Swift(XRT and UVOT), VLT/MUSE, AAVSO, CTIO:1.3m,
CTIO:1.5m,CXO}

\appendix
\section{Details of Individual Observations}
\label{app:obs_details}

Here, we present the observational details of each individual target with new ALMA data presented in this work. For each execution block, the number of antennas, the array configuration, and the on-target time are listed in Table \ref{tab:app_obs}.

\begin{deluxetable*}{lccccccc}
\setcounter{table}{0}
\renewcommand{\thetable}{A\arabic{table}}
\tablecaption{New Observations Presented in This Work\label{tab:app_obs}}
\tablewidth{0pt}
\tablehead{ \colhead{(1)} & \colhead{(2)} & \colhead{(3)} & \colhead{(4)} & \colhead{(5)} & \colhead{(6)} & \colhead{(7)}  \\
\colhead{System} & \colhead{ALMA Cycle} & \colhead{Set-up} & \colhead{Configuration} & \colhead{N} & \colhead{Observation Date}  & \colhead{On-Target Time} \\[-0.2cm]
\colhead{} & \colhead{} & \colhead{} & \colhead{} & \colhead{} & \colhead{}  & \colhead{(min)} }
%\colhead{($\arcsec\times\arcsec$
\startdata
\multicolumn{7}{c}{Dual AGN}\\
\hline
Mrk\,463 & Cycle\,10 & TM1 & C43-7 & 33 & Nov. 21, 2023 & 21.45 \\
& & TM1 & C43-7 & 43 & Nov. 22, 2023 & 38.08 \\
& & TM2 & C43-4 & 38 & Jan. 4, 2024 & 9.08 \\
& & TM2 & C43-4 & 47 & Jan. 7, 2024 & 9.10 \\
\hline
Mrk\,739 & Cycle\,10 & TM1 & C43-7 & 43 & Nov. 5, 2023 & 46.85 \\
& & TM1 & C43-7 & 44 & Nov. 21, 2023 & 46.87 \\
& & TM2 & C43-4 & 48 & Jan. 6, 2024 & 30.90 \\
\hline
UGC\,4211 & Cycle\,8 & TM1 & C43-8 & 46 & Oct. 24, 2021 & 31.45 \\
& & TM2 & C43-5 & 43 & Aug. 19, 2022 & 7.07 \\
\hline
ESO\,253-G003 & Cycle\,9 & TM1 & C43-9 & 43 & Jun. 25, 2023 & 45.37 \\
& & TM1 & C43-9 & 37 & Jun. 25, 2023 & 45.33 \\
& & TM1 & C43-9 & 40 & Jun. 26, 2023 & 45.33 \\
\hline\hline
\multicolumn{7}{c}{Single AGN}\\
\hline
Mrk\,975 & Cycle\,8 & TM1 & C43-8 & 36 & Oct. 24, 2021 & 39.00 \\
& & TM1 & C43-8 & 46 & Oct. 26, 2021 & 39.00 \\
& & TM1 & C43-8 & 46 & Oct. 26, 2021 & 39.02  \\
& & TM2 & C43-5 & 42 & Aug. 10, 2021 & 30.77 \\
\hline
NGC\,985 & Cycle\,8 & TM1 & C43-8 & 45 & Oct. 26, 2021 & 5.13 \\
& & TM1 & C43-8 & 45 & Oct. 27, 2021 & 5.13 \\
& & TM2 & C43-5 & 45 & Aug. 12, 2021 & 5.03 \\
\enddata
\tablecomments{Column (1): Observed system. Column (2): ALMA observation cycle number. Column (3): TM1 (long baseline) or TM2 (short baseline) set-up. Column (4): Array configuration. Column (5): Number of unflagged and usable antenna per execution. Column (6): Date of observation. Column (7): Total on-target observing time in minutes. NGC 6240 is not including in this table because we analyze archival ALMA data in this work. See \citet{treister20} for details of the observational set-up.}
\end{deluxetable*}

\section{KinMS MCMC Corner Plots}
\label{sec:corner}
\renewcommand\thefigure{\thesection.\arabic{figure}} 
\setcounter{figure}{0}  

Here, we present corner plots showing the 1- and 2-dimensional projections of the posterior probability distributions of the KinMS-fitted MCMC parameters. These plots were generated using the Python package \texttt{corner}. Note that the corner plots show the x- and y-coordinates of the rotating disk model's kinematic center in image pixel coordinates, whereas we report them in degrees in Table \ref{tab:mcmc}. 

\begin{figure*}
    \centering
    \includegraphics[width=\textwidth]{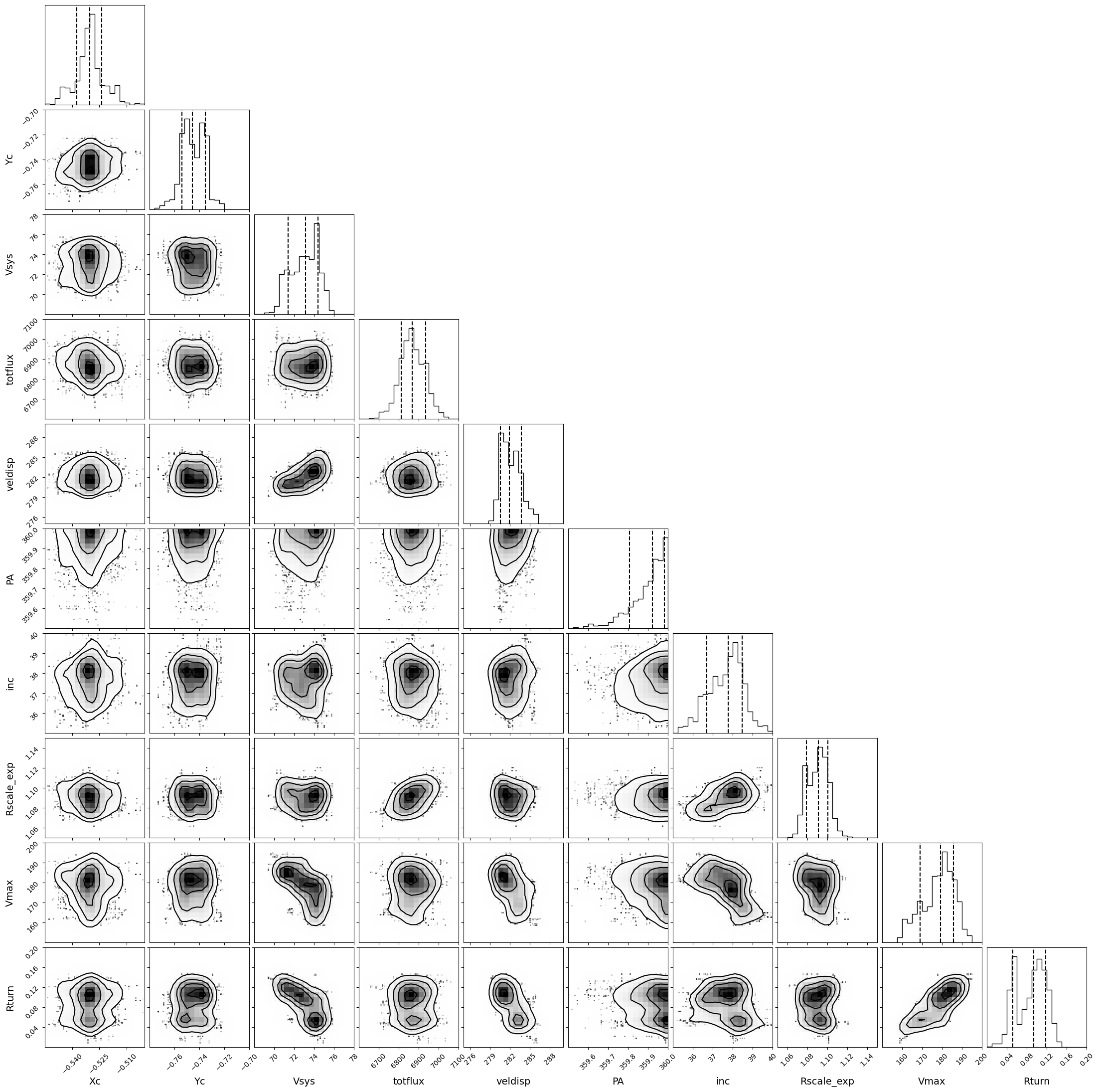}
    \caption{\textbf{NGC 6240} corner plot output from 30,000 iterations of our KinMS-implemented MCMC model fitting. }
\end{figure*}

%We acknowledge that a number of parameters (notably, $V_{\text{sys}}$,  $R_{\text{turn}}$, and  $V_{\text{max}}$) have double-peaked and/or non-Gaussian probability distributions. We emphasize, however, that the dynamic ranges of the plots are small, and the distinct peaks likely do not reflect physical features in the data. For example, the two peaks in the probability distribution of $R_{\text{turn}}$ differ by $\lesssim1$ beam-sized resolution element ($\sim0.06"$ or roughly 4 pixels). Similarly, the two peaks seen in $V_{\text{sys}}$ and $V_{\text{max}}$ are separated by less than the spectral resolution of the data cube ($\lesssim20$ km s$^{-1}$). The $1\sigma$ uncertainties of each parameter are reported in Table \ref{tab:mcmc}.

\begin{figure*}
    \centering
    \includegraphics[width=\textwidth]{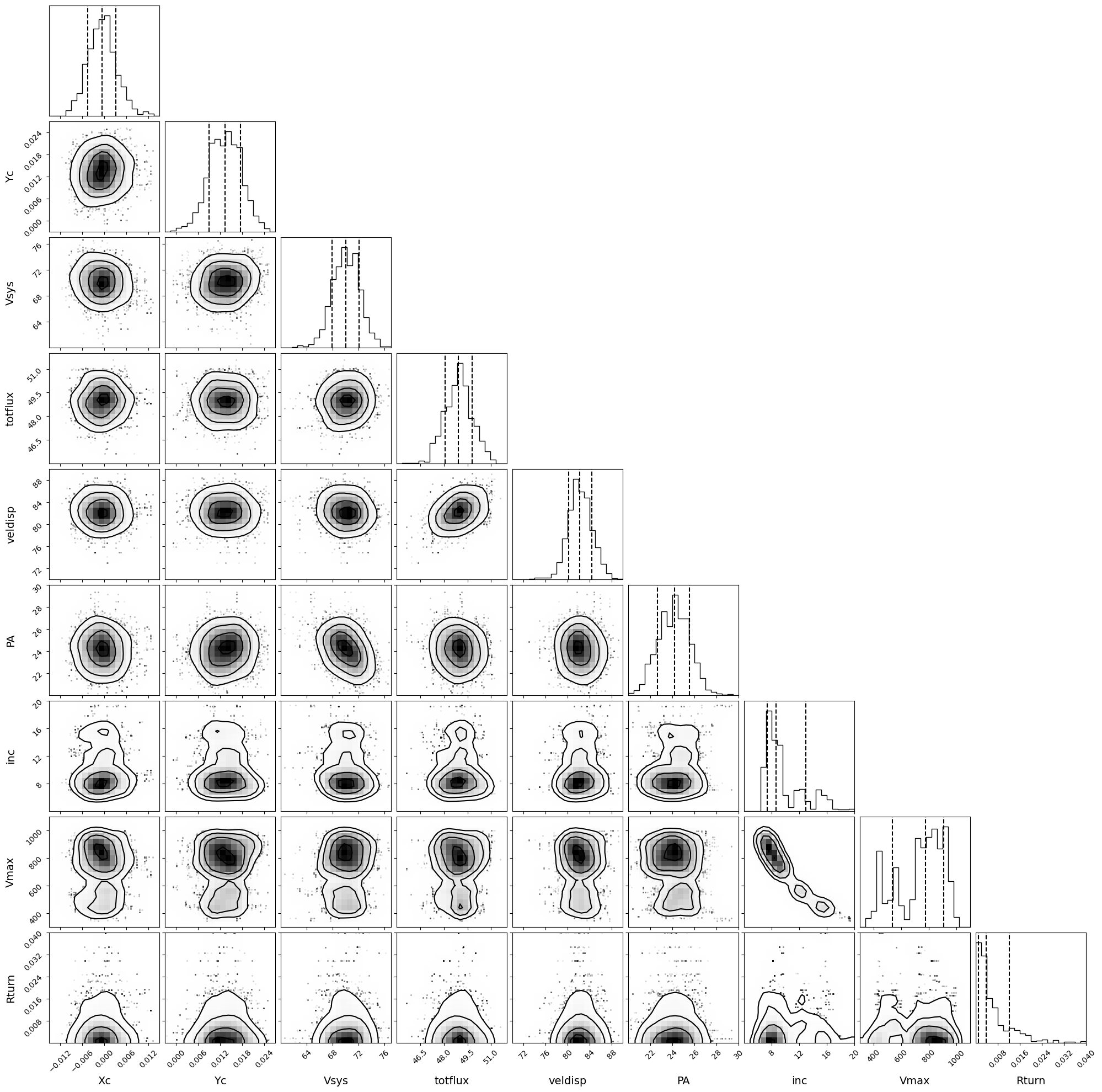}
    \caption{\textbf{Mrk 463E} corner plot output from 35,000 iterations of our KinMS-implemented MCMC model fitting. }
\end{figure*}

%We acknowledge that $V_{\text{max}}$ has a double-peaked probability distribution. This is likely due to the uncertainty in inclination, and the strong dependency between the two parameters.  Even so, the residual map of the best-fit model indicates a reasonable fit. The $1\sigma$ uncertainties of each parameter are reported in Table \ref{tab:mcmc}.

\begin{figure*}
    \centering
    \includegraphics[width=\textwidth]{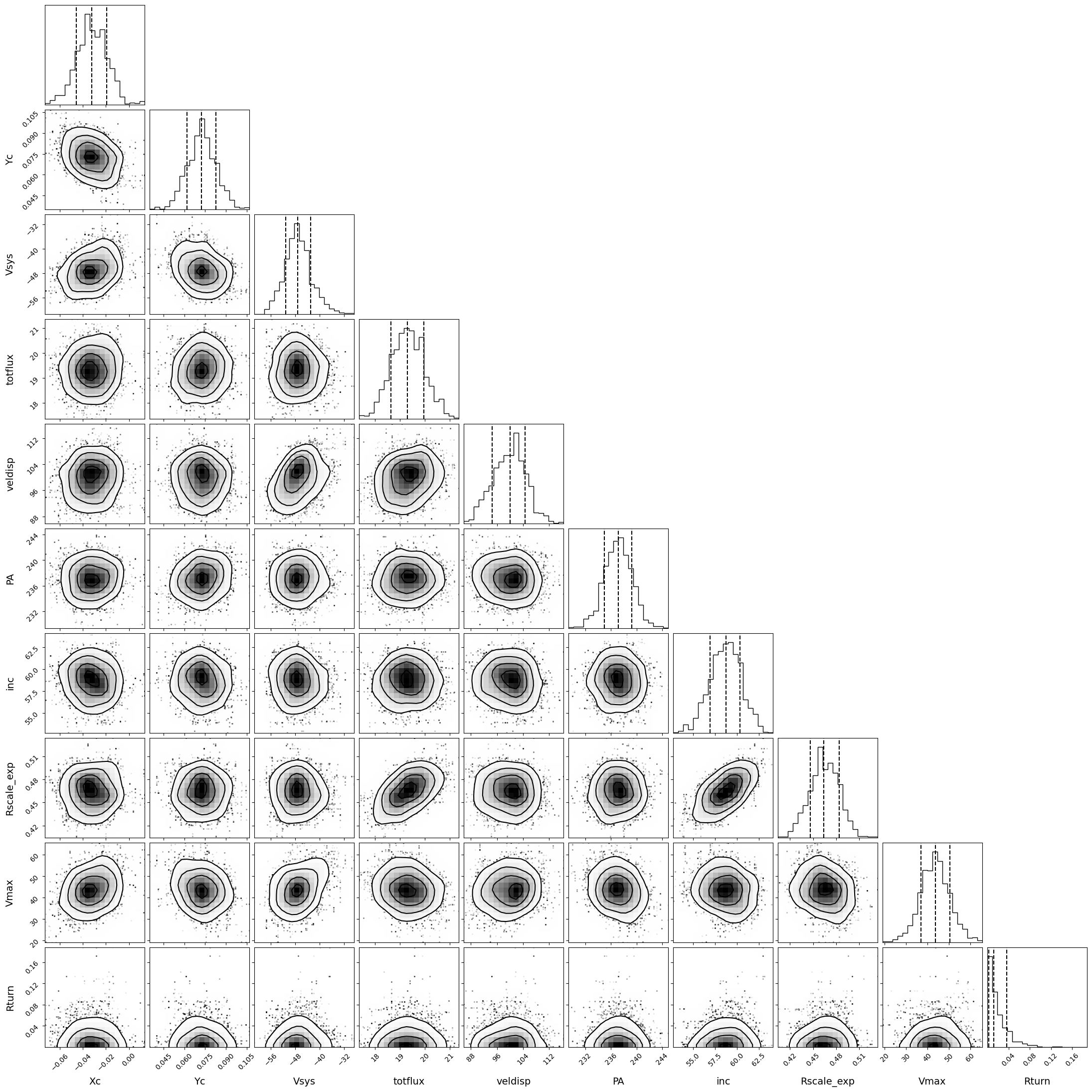}
    \caption{\textbf{Mrk 463W} corner plot output from 40,000 iterations of our KinMS-implemented MCMC model fitting. }
\end{figure*}

\begin{figure*}
    \centering
    \includegraphics[width=\textwidth]{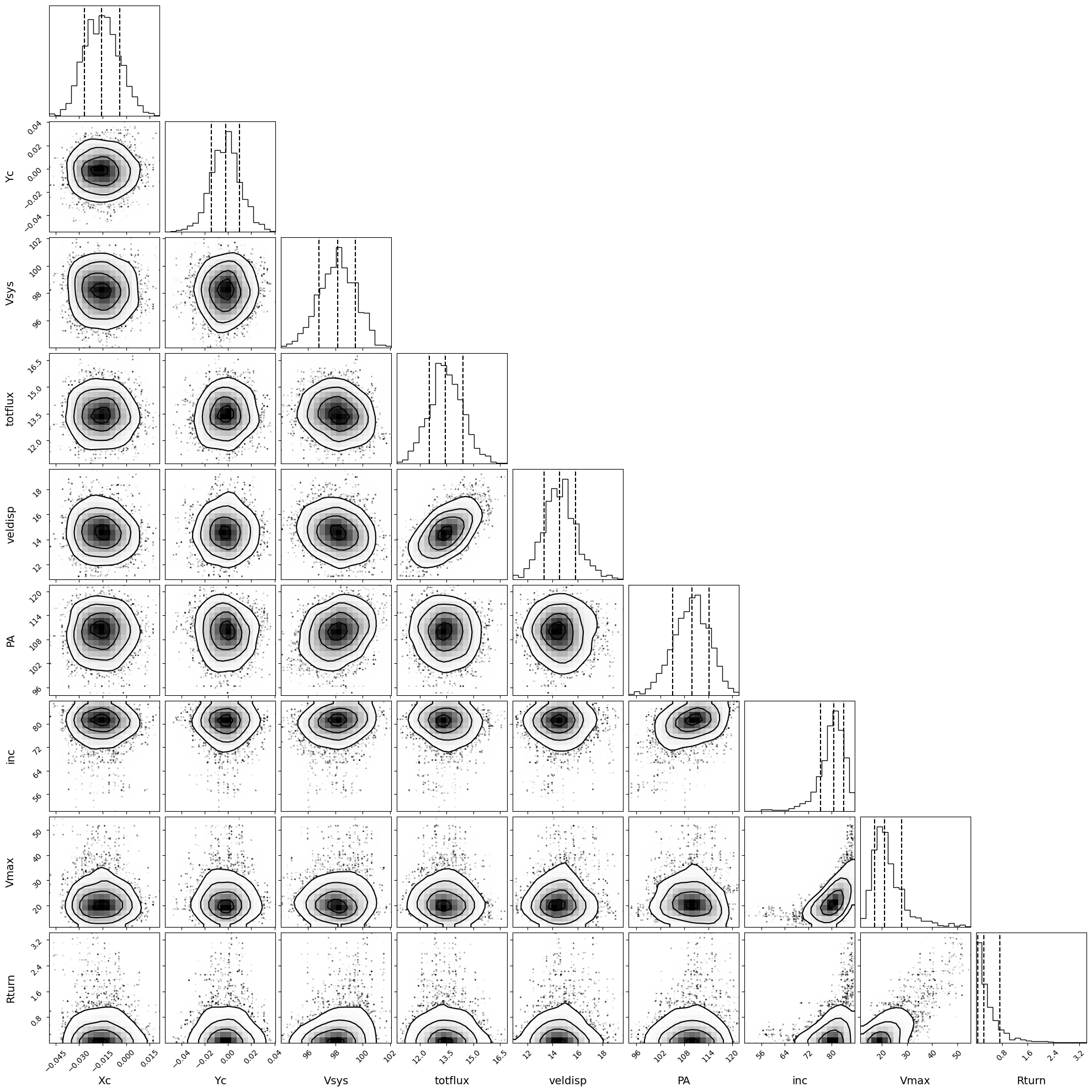}
    \caption{\textbf{Mrk 739E} corner plot output from 30,000 iterations of our KinMS-implemented MCMC model fitting.}
\end{figure*}

\begin{figure*}
    \centering
    \includegraphics[width=\textwidth]{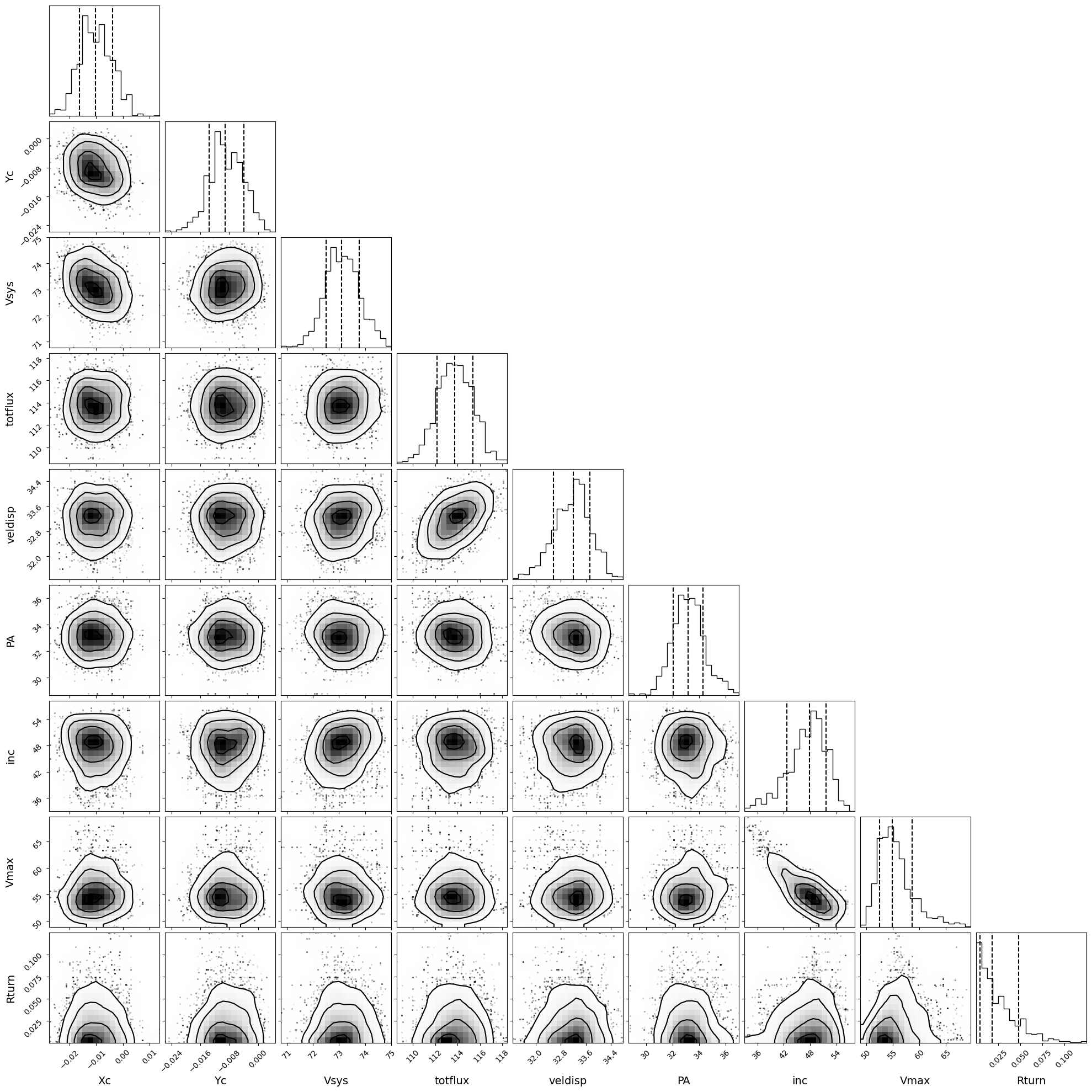}
    \caption{\textbf{Mrk 739W} corner plot output from 30,000 iterations of our KinMS-implemented MCMC model fitting. }
\end{figure*}

\begin{figure*}
    \centering
    \includegraphics[width=\textwidth]{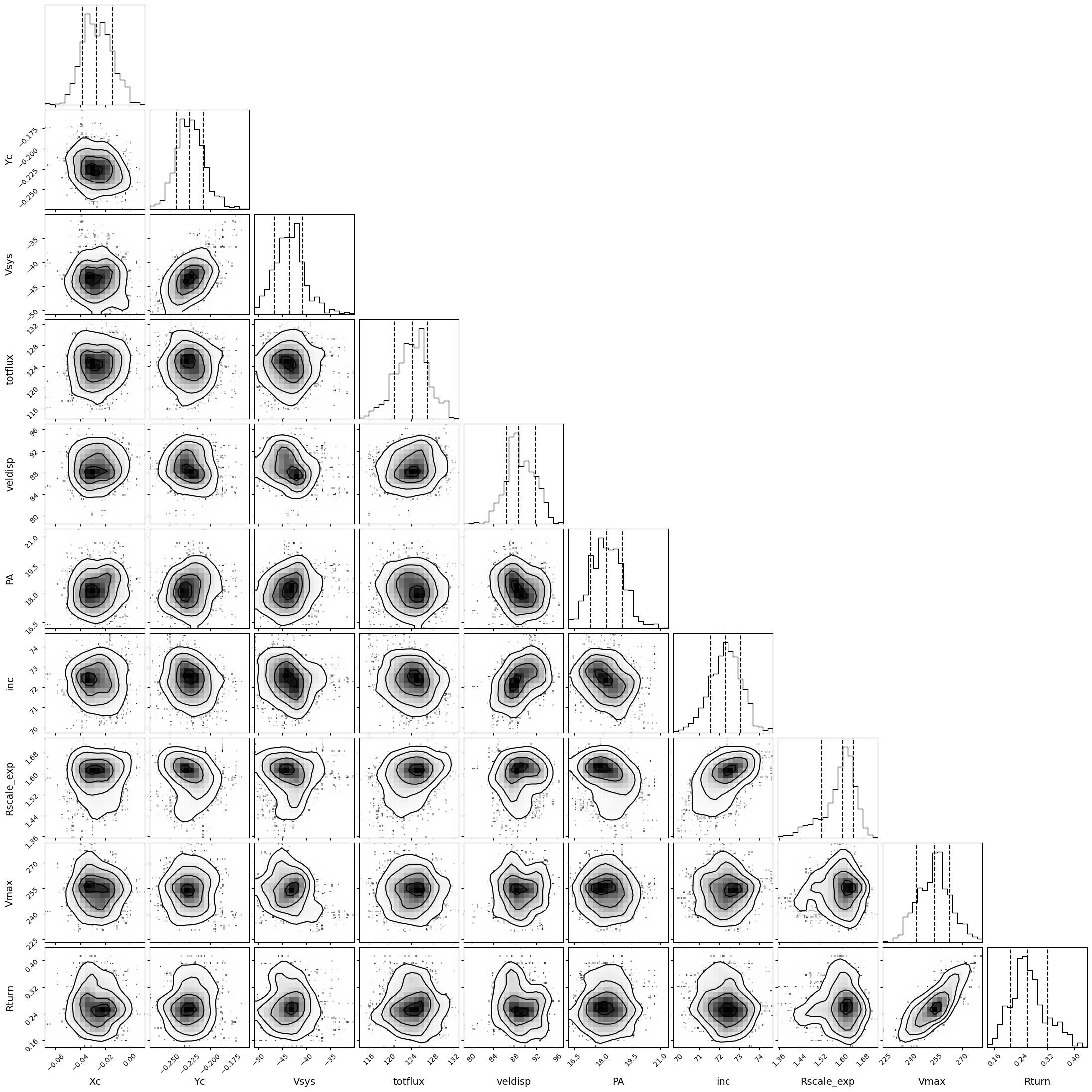}
    \caption{\textbf{UGC 4211} corner plot output from 30,000 iterations of our KinMS-implemented MCMC model fitting. }
\end{figure*}

\begin{figure*}
    \centering
    \includegraphics[width=\textwidth]{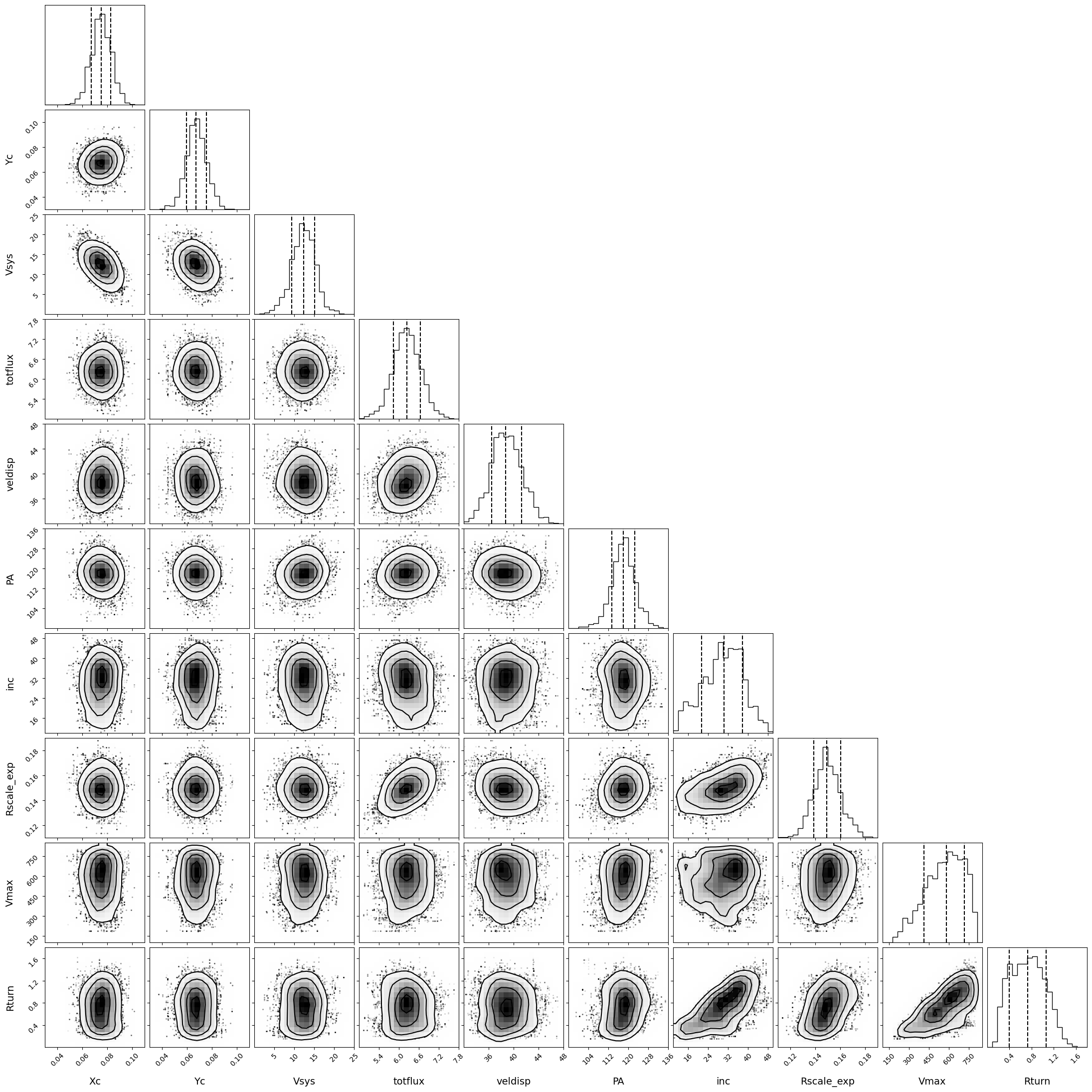}
    \caption{\textbf{ESO253-G003 NE} corner plot output from 65,000 iterations of our KinMS-implemented MCMC model fitting. }
\end{figure*}

\begin{figure*}
    \centering
    \includegraphics[width=\textwidth]{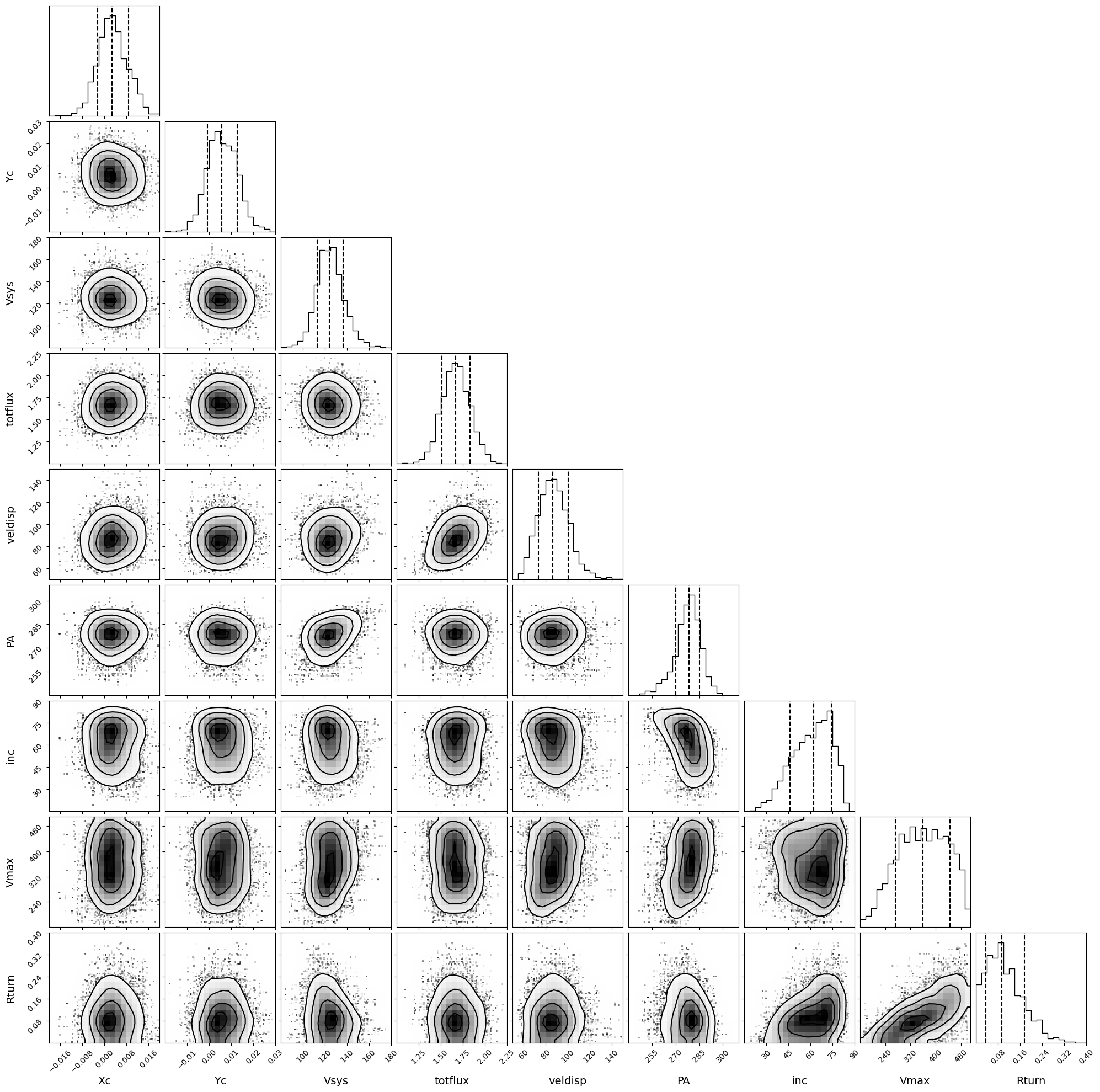}
    \caption{\textbf{ESO253-G003 SW} corner plot output from 60,000 iterations of our KinMS-implemented MCMC model fitting. }
\end{figure*}

%We acknowledge that the best-fit value for the $V_{\text{max}}$ parameter is highly uncertain. This likely results from the faint ($\sim3\sigma$) and compact morphology of the emission (see Figures \ref{fig:mom0_double} and \ref{fig:kinms_dual}), and may indicate a model-mismatch. Nevertheless, the SMBH SoI does not coincide with the majority of the observed $^{12}$CO(1-0) flux (see black circle in Figure \ref{fig:kinms_dual}), and we conclude that this uncertainty does have significant impacts on the results of our study. The $1\sigma$ uncertainties of each parameter are reported in Table \ref{tab:mcmc}.

\begin{figure*}
    \centering
    \includegraphics[width=\textwidth]{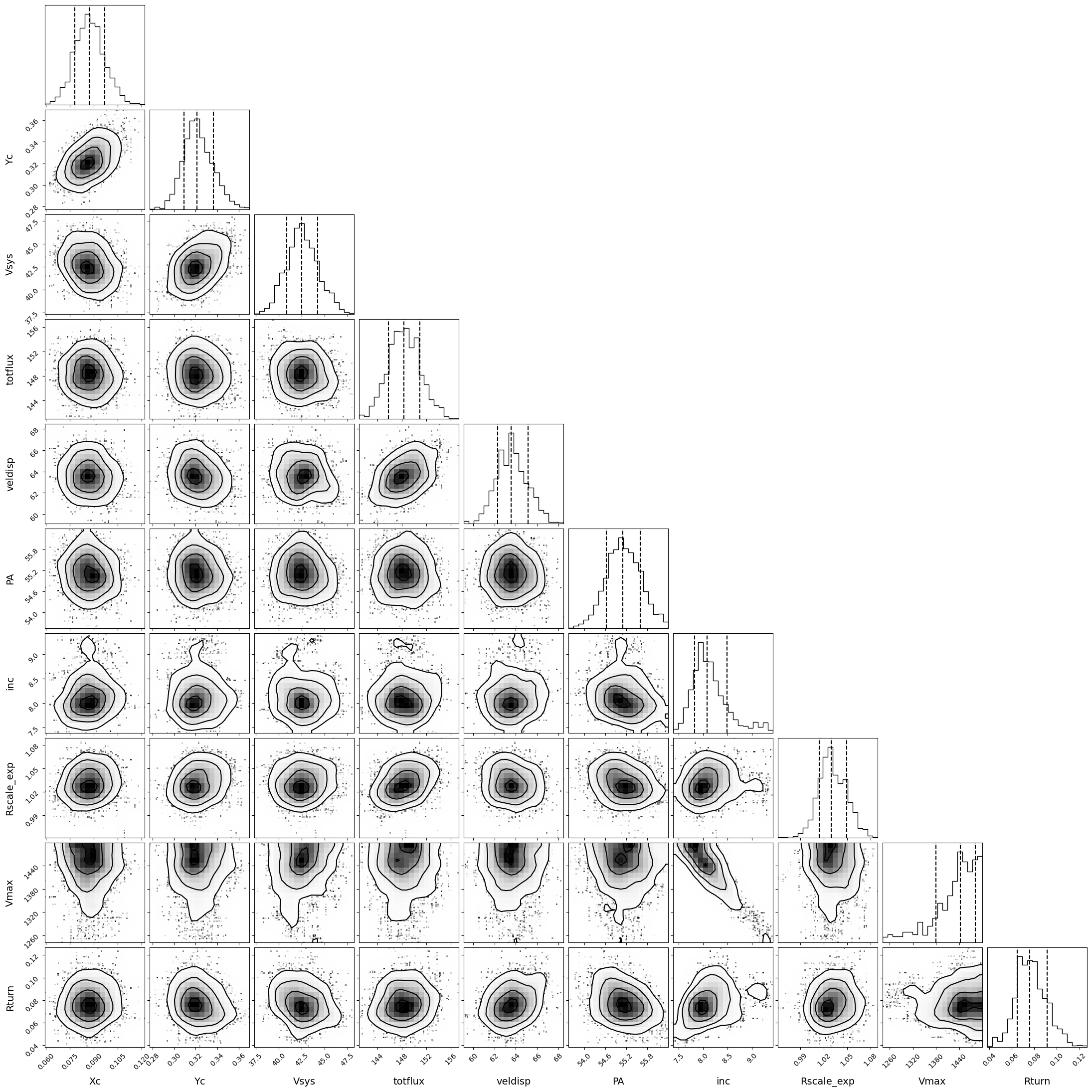}
    \caption{\textbf{Mrk 975} corner plot output from 40,000 iterations of our KinMS-implemented MCMC model fitting. $V_{\text{max}}$ was constrained to $V_{\text{max}}\leq1500$ km s$^{-1}$ (the maximum velocity range of the inputted data cube) to ensure physically meaningful results. }
\end{figure*}

%$V_{\text{max}}$ was constrained to $V_{\text{max}}<1500$ km s$^{-1}$ (the maximum velocity of the inputted data cube) to ensure physically meaningful results. The $1\sigma$ uncertainties of each parameter are reported in Table \ref{tab:mcmc}.

%\software{astropy \citep{2013A&A...558A..33A,2018AJ....156..123A},  
%          Cloudy \citep{2013RMxAA..49..137F}, 
%          Source Extractor \citep{1996A&AS..117..393B}
%          }

\bibliography{bib.bib}{}
\bibliographystyle{aasjournal}

\end{document}